\documentclass[12pt,preprint]{aastex}

\shorttitle{Long-term Evolution of a Black Hole}
\shortauthors{Fujita}

\begin{document}

\title{Long-Term Evolution of and X-ray Emission from a Recoiling
Supermassive Black Hole in a Disk Galaxy}

\author{Yutaka Fujita}
\affil{Department of Earth and Space Science, Graduate School
of Science, Osaka University, \\
1-1 Machikaneyama-cho, Toyonaka, Osaka
560-0043, Japan}

\begin{abstract}
Recent numerical relativity simulations have shown that the emission of
gravitational waves at the merger of two black holes gives a recoil kick
to the final black hole. We follow the orbits of a recoiling
supermassive black hole (SMBH) in a fixed background potential of a disk
galaxy including the effect of dynamical friction. If the recoil
velocity of the SMBH is smaller than the escape velocity of the galaxy,
the SMBH moves around in the potential along a complex trajectory before
it spirals into the galactic center through dynamical friction. We
consider the accretion of gas onto the SMBH from the surrounding ISM and
estimate the X-ray luminosity of the SMBH. We find that it can be larger
than $3\times 10^{39}\rm\: erg^{-1}$ or the typical X-ray luminosity of
ultra-luminous X-ray sources, when the SMBH passes the galactic disk. In
particular, the luminosity could exceed $\sim 10^{46}\:\rm erg\:
s^{-1}$, if the SMBH is ejected into the galactic disk. The average
luminosity gradually increases as the SMBH spirals into the galactic
center. We also estimate the probability of finding recoiling SMBHs with
X-ray luminosities of $>3\times 10^{39}\rm\: erg^{-1}$ in a disk galaxy.
\end{abstract}

\keywords{black hole physics --- ISM: general --- galaxies: nuclei  ---
X-rays: general}

\section{Introduction}
\label{sec:intro}

Thanks to recent breakthroughs in numerical relativity, it has been
shown that the loss of linear momentum radiated away in the form of
gravitational waves induces a large recoil velocity of the merged binary
black hole. This would have been happened for supermassive black holes
(SMBHs) at the centers of galaxies, if two SMBHs coalesce after a major
galaxy merger. Since the maximum velocity would reach $\sim 4000\rm\:
km\: s^{-1}$ \citep[e.g.][]{gon07,cam07}, the SMBH could escape from its
host galaxy. However, if the recoil velocity is a little smaller than
the escape velocity of the galaxy, the SMBH would orbit in the potential
well of the galaxy for a long time. The identification of observational
signatures of such recoiling SMBHs is important for studies about the
growth of black holes as well as the general relativity.

In addition to direct detection of gravitational waves, a number of
ideas have been proposed for detection of observational signatures of
recoiling black holes through electromagnetic waves. \citet{kap76,kap83}
indicated that stellar captures can lead to the formation of an
accretion disk-star system about the SMBH, and that the emission from
the SMBH could be observable. \citet{mad04} and \citet{loe07} argued
that a recoiling SMBH would be observed as an off-nuclear quasar until
the gas carried by the SMBH is depleted, although \citet*{bon07} found
no convincing evidence for recoiling SMBHs carrying accretion disks in
the Sloan Digital Sky Survey data. \citet{mer04} indicated that the
displacement of a recoiling SMBH transfers energy to the stars in the
galactic nucleus and converts a steep density cusp into a
core. \citet{vol07} discussed the influence of the merger and ejection
of SMBHs from the galactic centers on the relation of the black hole
mass and the velocity dispersion of the galaxy. \citet{gua08} indicated
that helical radio structures could be observed around a recoiling SMBH
because of the oscillation of the SMBH in the core of the host
galaxy. \citet*{lip08} showed that prompt shocks are created in the gas
disk around a recoiling SMBH and that the shocks could result in an
afterglow, and the luminosity and characteristic photon energy increases
with time. \citet{kor08} discussed that the SMBH displacement may give
rise to observable non-axisymmetries in the morphology and dynamics of
the stellar and gaseous disk of the host galaxy. \citet{del08} examined
the influence of a runaway SMBH passage through intergalactic medium,
and indicated that the SMBH is able to ignite star formation efficiently
in the wake of its trajectory. \citet{kom08} indicated that a recoiling
SMBH carries stars, and the electromagnetic flares from the stars that
are tidally disrupted by the SMBH would be observable.

Recently, \citet{ble08} calculated the trajectory of a SMBH ejected in a
smooth background potential that includes both a stellar bulge and a
gaseous disk \citep*[see also][]{vic07}, and estimated the gas accretion
rate onto the SMBH as a function of time. \citet[][hereafter
Paper~I]{fuj08} also calculated the trajectory of a SMBH ejected in a
realistic background potential of a disk galaxy. We calculated the
accretion rate of gas onto the SMBH from the interstellar medium (ISM)
in the galactic disk, and estimated the X-ray luminosity based on a
model of a radiatively inefficient accretion flow
\citep[RIAF;][]{nar95}. We showed that the luminosity of the SMBH can be
comparable to or even larger than those of ultra-luminous X-ray sources
(ULXs) observed in galaxies \citep[$L_X\ga 3\times 10^{39}\rm\: erg\:
s^{-1}$;][]{col99,mak00,mus04}.

However, in that study, the effect of dynamical friction was not
explicitly included. If dynamical friction is effective, the SMBH
gradually settles down to the galactic center. Since the accretion rate
depends on the density of the surrounding ISM (see
equation~[\ref{eq:dotm}]), we expect that the X-ray luminosity increases
accordingly. In this paper, we study the long-term orbital and
luminosity evolution of a recoiling SMBH in a disk galaxy, considering
the effect of dynamical friction. This paper is organized as
follows. Our models and choice of parameters are outlined in
\S~\ref{sec:models}. The results of calculations are presented in
\S~\ref{sec:results} and discussed in \S~\ref{sec:discussion}. Finally,
in \S~\ref{sec:conclusion}, we present our conclusions.

\section{Models}
\label{sec:models}

We consider SMBH mergers in a normal disk galaxy for the sake of
simplicity, although the galaxy interaction and merger supposedly affect
the original disk. However, the process of the settling of SMBHs between
the galaxy merger and the set-in of effective emission of gravitational
waves from the SMBHs, which leads to the merger of the SMBHs, has not
been understood \citep*[e.g.][]{beg80,iwa06}.  If the time-scale of the
SMBH merger is long, the galaxy may be significantly relaxed when the
recoil occurs \citep{ble08}. Anyway, in order to follow the galaxy and
SMBH mergers self-consistently, we would need ultra-high resolution
simulations of galaxy mergers that can resolve the settling of SMBHs in
the core of the merged galaxy.

The model of a disk galaxy is the same as that in Paper~I.  The galaxy
potential consists of three components, which are a \citet{miy75} disk,
Hernquist spheroid \citep{her90}, and logarithmic halo \citep{bin08}:
\begin{equation}
 \Phi_{\rm disk}=-\frac{G M_{\rm dist}}
{\sqrt{R^2+(a+\sqrt{z^2+b^2})^2}}\:,
\end{equation}
\begin{equation}
\Phi_{\rm sphere}=-\frac{G M_{\rm sphere}}{r+c}\:,
\end{equation}
\begin{equation}
\Phi_{\rm halo}=\frac{1}{2}v_{\rm halo}^2
\ln\left[R^2+\left(\frac{z}{q}\right)^2+d^2\right]\:,
\end{equation}
where $R$ ($=\sqrt{x^2+y^2}$) and $z$ are cylindrical coordinates
aligned with the galactic disk, and $r=\sqrt{R^2+z^2}$. The parameters
are the ones of the Galaxy. We take $M_{\rm disk}=1.0\times 10^{11}\:
M_\sun$, $M_{\rm sphere}=3.4\times 10^{10}\: M_\sun$, $a=6.5$~kpc,
$b=0.26$~kpc, $c=0.7$~kpc, $d=13$~kpc, and $q=0.9$; $v_{\rm halo}$ is
determined so that the circulation velocity for the total potential,
$\Phi=\Phi_{\rm disk}+\Phi_{\rm sphere}+\Phi_{\rm halo}$, is $220\rm\:
km\: s^{-1}$ at $R=7$~kpc \citep*[see][]{law05}. Contrary to Paper~I, we
include dynamical friction in the equation of motion for the SMBH
\citep{cha43,bin08}:
\begin{equation}
\label{eq:motion}
 \mbox{\boldmath{$\dot{v}$}}=-\nabla\Phi-4\pi G^2 m_{\rm
 BH}\sum_{i}\rho_i I(X_i)\ln\Lambda_i
 \frac{\mbox{\boldmath{$v$}}_{{\rm rel}, i}}{v_{{\rm rel},i}^3}\:,
\end{equation}
where $\mbox{\boldmath{$v$}}=(v_x, v_y, v_z)$ is the velocity of the SMBH,
$\ln\Lambda_i$ is the Coulomb logarithm, and the suffix $i$ refers to
disk, sphere, or halo. The relative velocities are defined as
$\mbox{\boldmath{$v$}}_{\rm rel,
disk}=\mbox{\boldmath{$v$}}-\mbox{\boldmath{$v$}}_{\rm cir,disk}$,
$\mbox{\boldmath{$v$}}_{\rm rel, sphere}=\mbox{\boldmath{$v$}}$, and
$\mbox{\boldmath{$v$}}_{\rm rel, halo}=\mbox{\boldmath{$v$}}$, where
$\mbox{\boldmath{$v$}}_{\rm cir, disk}$ is the circulation velocity of
the disk, which is given by
\begin{equation}
 v_{\rm cir, disk}^2=\left. R\frac{\partial \Phi}{\partial R}\right|_{z=0}
\end{equation}
\citep{bin08}. The densities are given by
\begin{equation}
 \rho_{\rm disk}=\left(\frac{b^2 M_{\rm disk}}{4\pi}\right)
\frac{a R^2+(a+3\sqrt{z^2+b^2})(a+\sqrt{z^2+b^2})^2}
{[R^2+(a+\sqrt{z^2+b^2})^2]^{5/2}(z^2+b^2)^{3/2}}\:,
\end{equation}
\begin{equation}
 \rho_{\rm sphere}=\frac{M_{\rm sphere}}{2\pi}\frac{c}{r(c+r)^3}\:,
\end{equation}
\begin{equation}
 \rho_{\rm halo}=\frac{v_{\rm halo}^2}{4\pi G
  q^2}\frac{(2q^2+1)d^2+R^2+(2-q^{-2})z^2}{(d^2+R^2+z^2 q^{-2})^2}\:,
\end{equation}
\citep{miy75,her90,bin08}. We assume that part of the disk consists of
the ISM; its density is represented by $\rho_{\rm ISM}=f_{\rm
ISM}\rho_{\rm disk}$. For most models we studied, we take $f_{\rm
ISM}=0.2$, which is based on the observations of the Galaxy
\citep[e.g.][]{bin08}. In equation~(\ref{eq:motion}), the factor
$I(X_i)$ is given by
\begin{equation}
 I(X_i)={\rm erf}(X_i)-\frac{2X_i}{\sqrt{\pi}}e^{-X_i^2}
\end{equation}
where $X_i=v_{{\rm rel},i}/(\sqrt{2}\sigma_i)$ and $\sigma_i$ is the
one-dimensional velocity dispersion. For the disk velocity dispersion,
we set $\sigma_{\rm disk}\propto \rho_{\rm disk}(x,y,z=0)$ \citep{lew89}
and fix the normalization by assuming that the disk has a Toomre
Q-parameter of 1.5 at $R=7$~kpc \citep{vel99}. For the spheroid and 
halo, we use the common $I(X_i)$ and the velocity dispersion is
\begin{equation}
 \sigma_{\rm sphere-halo}
=\frac{(v_{\rm cir, sphere}^2+v_{\rm cir, halo}^2)^{1/2}}{\sqrt{2}}
\end{equation}
where $v_{\rm cir, sphere}$ and $v_{\rm cir, halo}$ are the circulation
velocities of the spheroid and halo, respectively \citep{tay01}.

Chandrasekhar's formula for the dynamical friction force (the second
term in the right side of equation~[\ref{eq:motion}]) was derived
assuming an infinite, homogeneous, and unchanging background. It is
obviously not true for a SMBH that is kicked out of the galactic center
and orbiting in the disk galaxy. However, $N$-body simulations have
shown that it can be applied to various cases if one chooses the Coulomb
logarithm appropriately. For a SMBH ejected from the center of a
spherically symmetric galaxy, \citet{gua08} showed that
$2\la\ln\Lambda\la 3$ is appropriate. Thus, we set $\ln\Lambda_{\rm
sphere}=\ln\Lambda_{\rm halo}=2.5$. For a disk galaxy, such $N$-body
simulations have not been performed as far as we know. Instead of a
SMBH, the evaluation of Chandrasekhar's formula has been made for a
dwarf galaxy infalling to a more massive disk galaxy. \citet{tay01}
showed that $\ln\Lambda_{\rm disk}=0.5$ is appropriate. The small value
of $\ln\Lambda_{\rm disk}$ probably reflects the small ratio of the disk
scale-height and the size of a dwarf galaxy, which corresponds to the
ratio of maximum and minium effective impact parameters of particles
that contribute to the friction force \citep{tay01}. However, since a
SMBH is a point source, we first assume that $\ln\Lambda_{\rm
disk}=2.5$, which is the same as $\ln\Lambda_{\rm sphere}$ and
$\ln\Lambda_{\rm halo}$, and then change the value to see the influence
of the uncertainty on results.

The accretion of the surrounding gas onto an isolated black hole has
been studied by several authors \citep*[][and references
therein]{fuj98,ago02,mii05,map06}. Most of the previous studies focused
on stellar mass ($\sim 10\: M_\sun$) or intermediate mass black holes
(IMBHs; $\sim 10^3\: M_\sun$). In this study, we consider the accretion
on a recoiling SMBH.

The accretion rate of the ISM onto the SMBH is given by the Bondi-Hoyle
accretion \citep{bon52}:
\begin{equation}
\label{eq:dotm}
 \dot{m}=2.5\pi G^2
\frac{m_{\rm BH}^2\rho_{\rm ISM}}{(c_s^2+v_{\rm rel, disk}^2)^{3/2}}\:,
\end{equation}
where $m_{\rm BH}$ is the mass of the black hole, and $c_s$ ($=10\rm\:
km\: s^{-1}$) is the sound velocity of the ISM.
The X-ray luminosity of the black hole is given by
\begin{equation}
\label{eq:Lx}
 L_{\rm X}=\eta \dot{m}c^2\:,
\end{equation}
where $\eta$ is the efficiency.  Since the accretion rate is relatively
small for the mass of the black hole, the accretion flow would be a RIAF
\citep[e.g.][]{ich77,nar94,abr95,yua04}. In this case, the efficiency
follows $\eta\propto\dot{m}$ for $L_{\rm X}\la 0.1 L_{\rm Edd}$, where
$L_{\rm Edd}$ is the Eddington luminosity
\citep*[e.g.][]{kat98}. Therefore, we assume that $\eta=\eta_{\rm Edd}$
for $\dot{m}> 0.1 \dot{m}_{\rm Edd}$ and $\eta = \eta_{\rm
Edd}\dot{m}/(0.1\dot{m}_{\rm Edd})$ for $\dot{m}< 0.1 \dot{m}_{\rm
Edd}$, where $\dot{m}_{\rm Edd}=L_{\rm Edd}/(c^2 \eta_{\rm Edd})$
\citep{mii05}. We assume that $\eta_{\rm Edd}=0.1$.

We solve equation~(\ref{eq:motion}) with Mathematica 6.0 using a command
NDSolve. The algorism of the integration is automatically chosen
\footnote{http://support.wolfram.com/mathematica/mathematics/numerics/ndsolvereferences.en.html}. We
have confirmed that the fractional energy error that arises in the
integration of an orbit per cycle is $\lesssim 10^{-6}$.

\section{Results}
\label{sec:results}

The SMBH is placed at the center of the galaxy at $t=0$. The SMBH is
ejected on the $x$-$z$ plane at $t=0$. The parameters of our models
($m_{\rm BH}$, $v_0$, $\ln\Lambda_{\rm disk}$, and $f_{\rm ISM}$) are
shown in Table~\ref{tab:par}. In this section, we consider models in
which the SMBH does neither fall into the galactic center too quickly
through dynamical friction nor escape from the galaxy (models~A1--C4 in
Table~\ref{tab:par}). The mass of the SMBH is $3\times 10^6$--$3\times
10^7\: M_\sun$, which is comparable to or somewhat larger than that of
the SMBH at the center of the Galaxy \citep[$\sim 3.7\times 10^{6}\:
M_\sun$;][]{sch02,ghe05}. The direction of the ejection changes from
$\theta=0\arcdeg$ to $90\arcdeg$, where $\theta=0\arcdeg$ corresponds to
the $z$-axis. The initial velocity of the SMBH is $v_0=500$--$800\:\rm
km\: s^{-1}$.

\begin{figure}
\epsscale{1.0} \plottwo{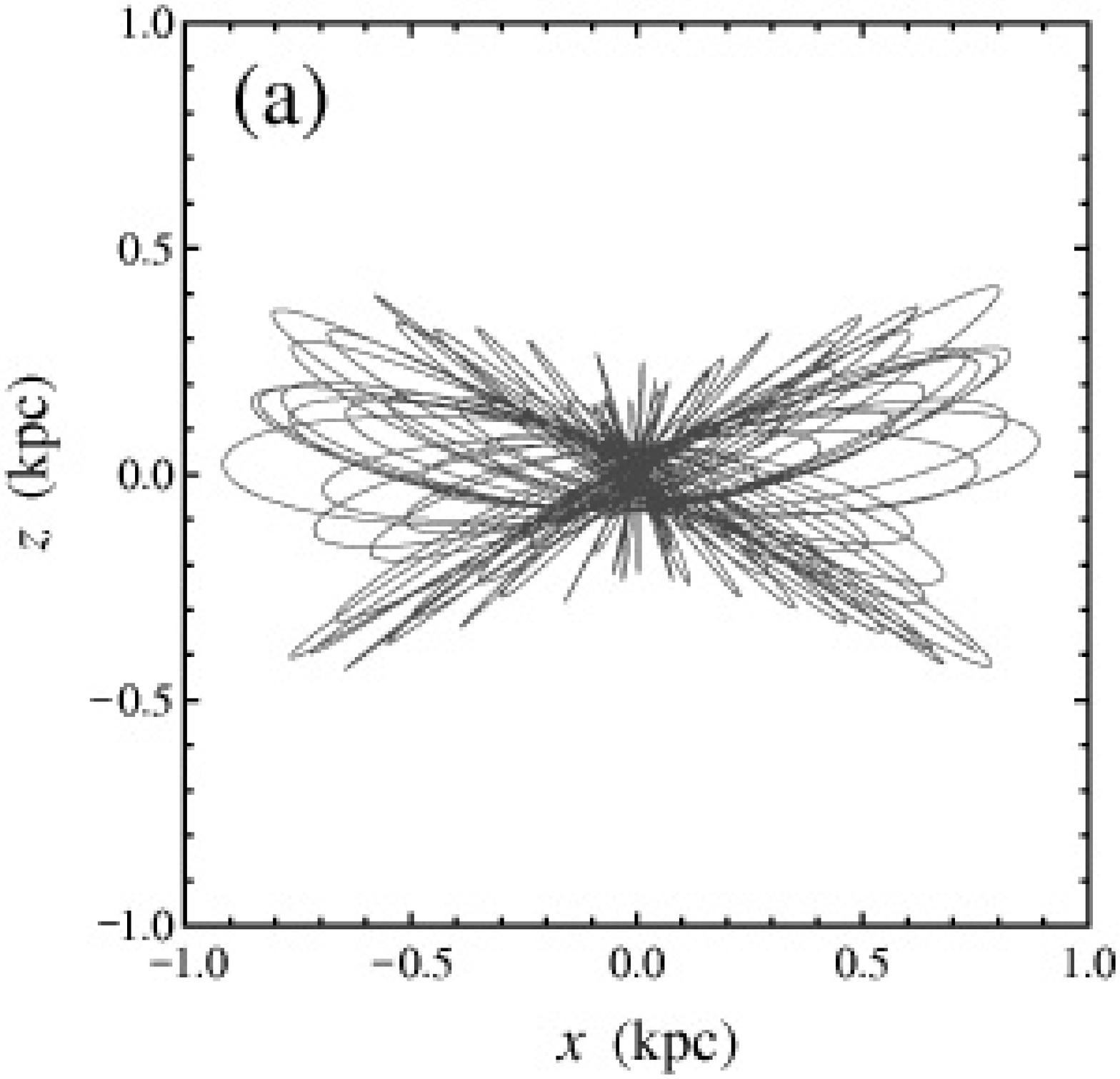}{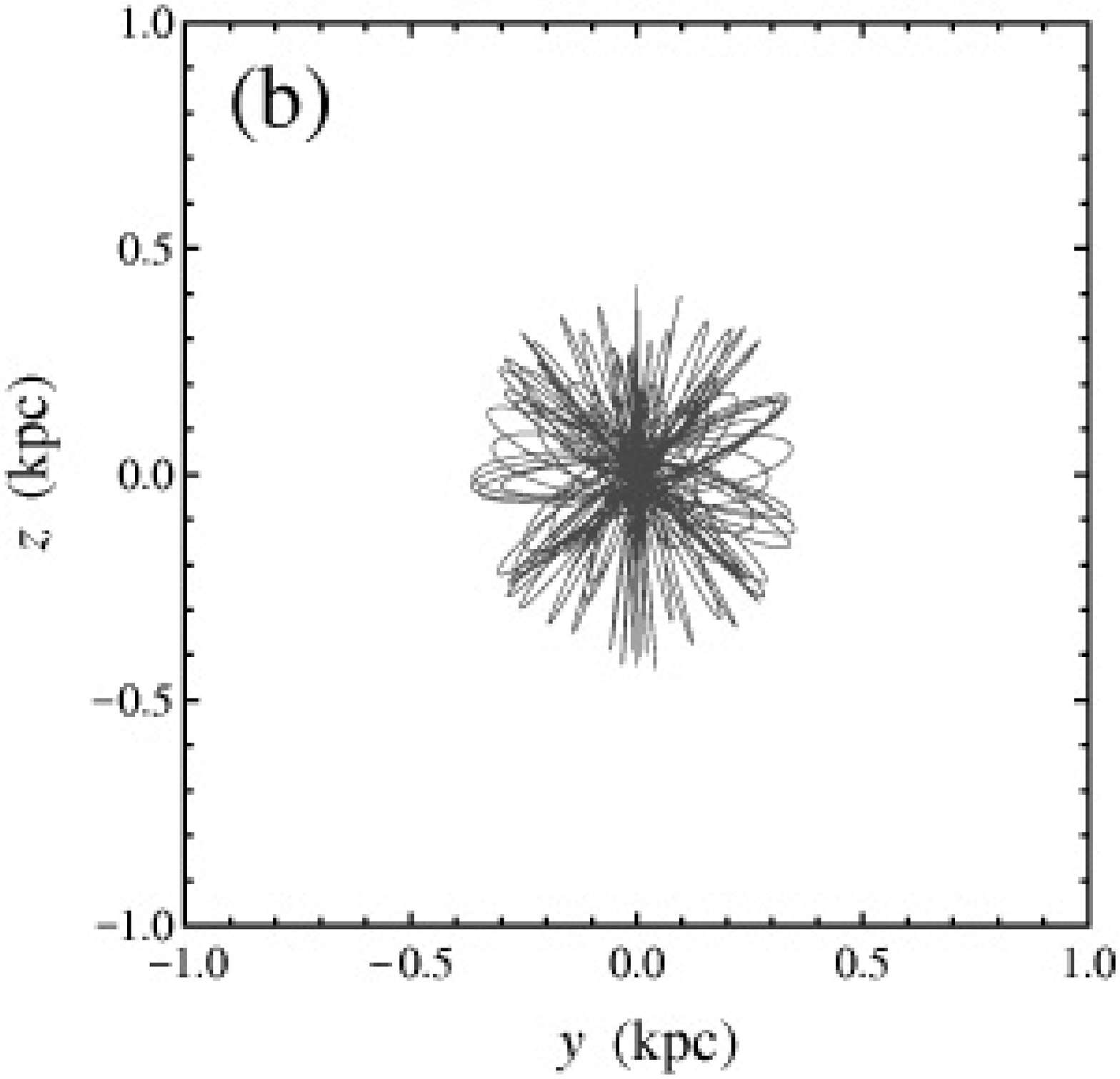} \caption{The trajectory of the
SMBH for model~A1 for $0<t<t_{\rm df}$ when
$\theta=60\arcdeg$. \label{fig:orbitA1}}
\end{figure}

\begin{figure}
\epsscale{1.0} \plottwo{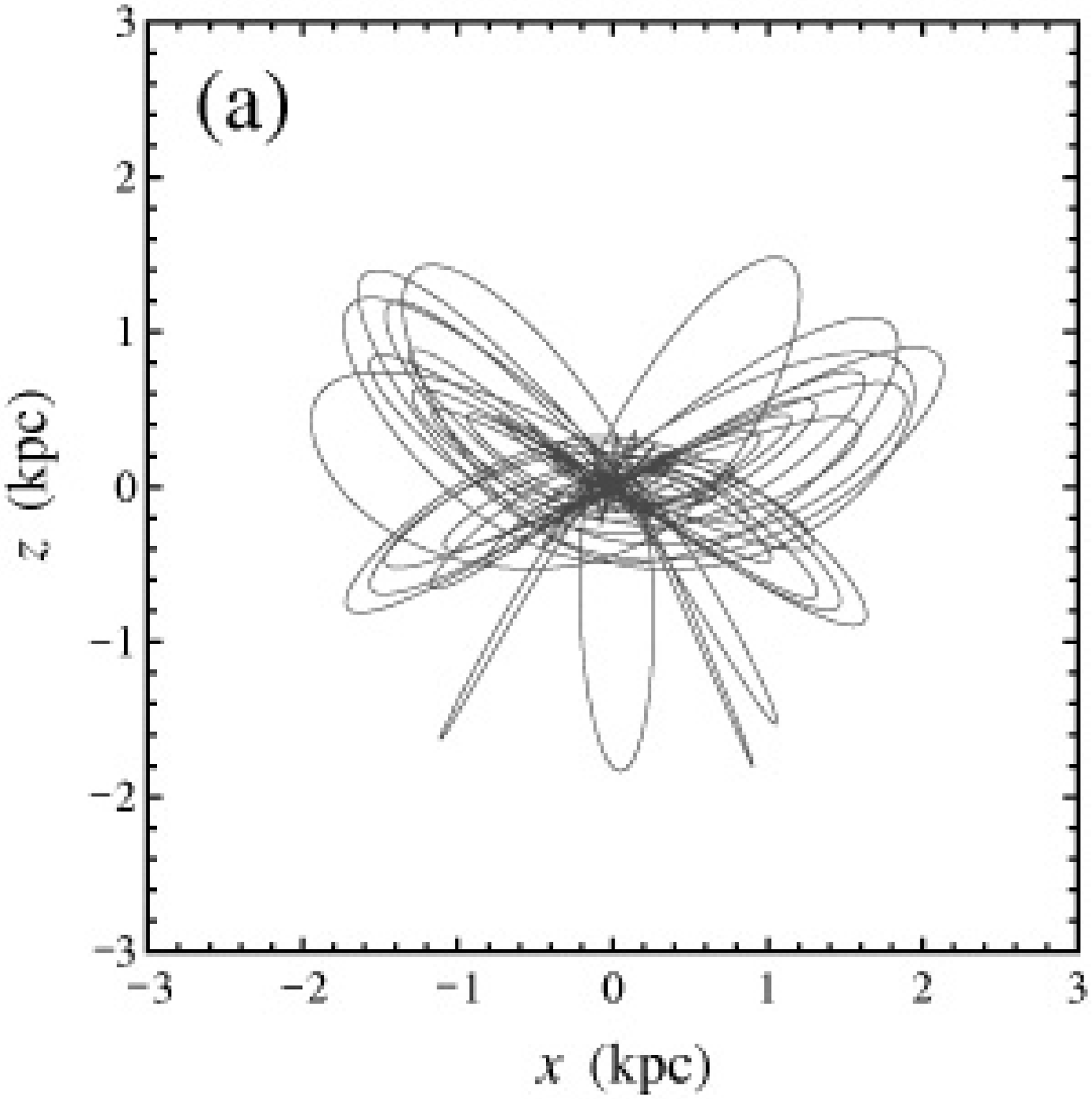}{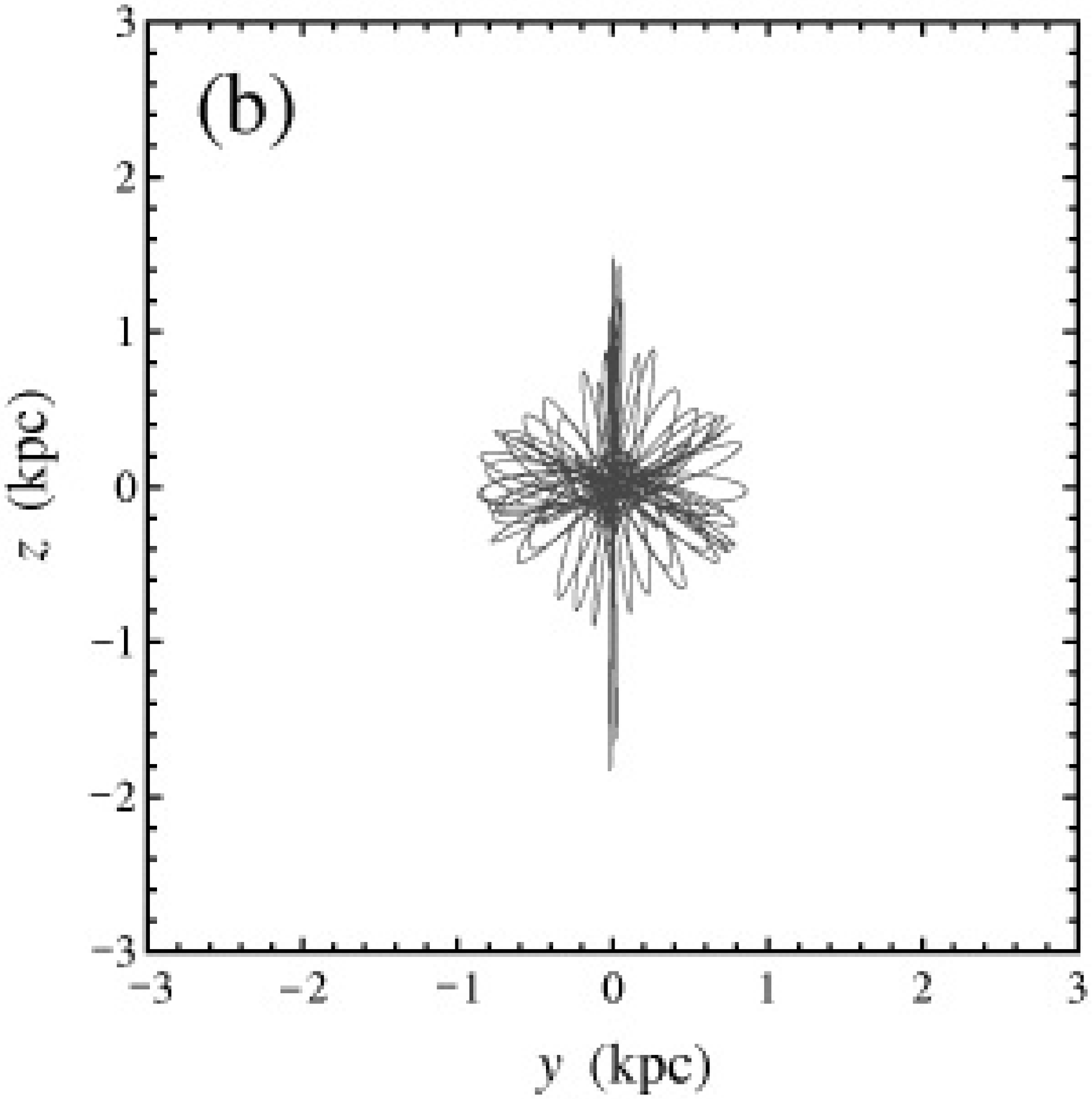} \caption{Same as
 Figure~\ref{fig:orbitA1} but for model~B2. \label{fig:orbitB2}}
\end{figure}

\begin{figure}
\epsscale{1.0} \plottwo{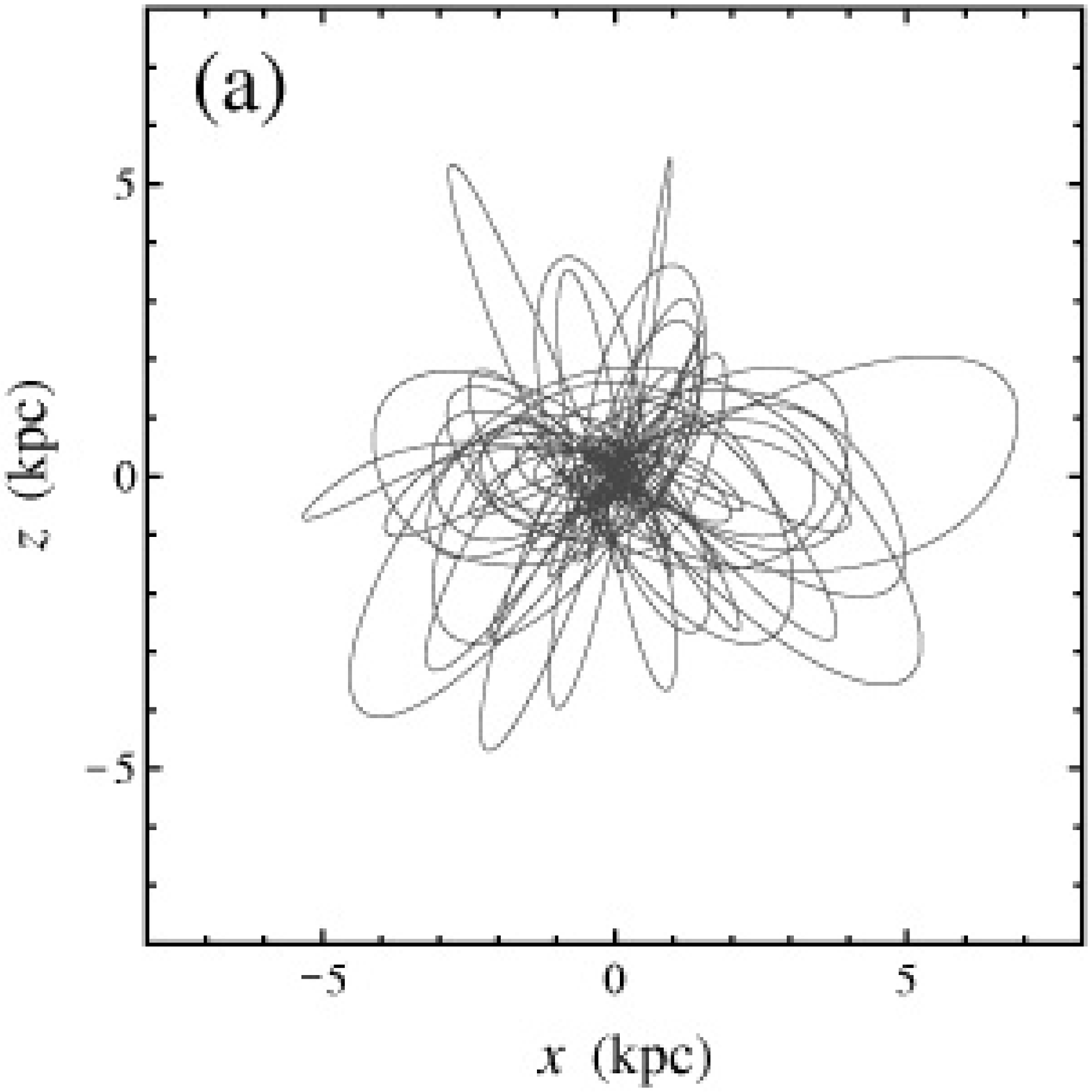}{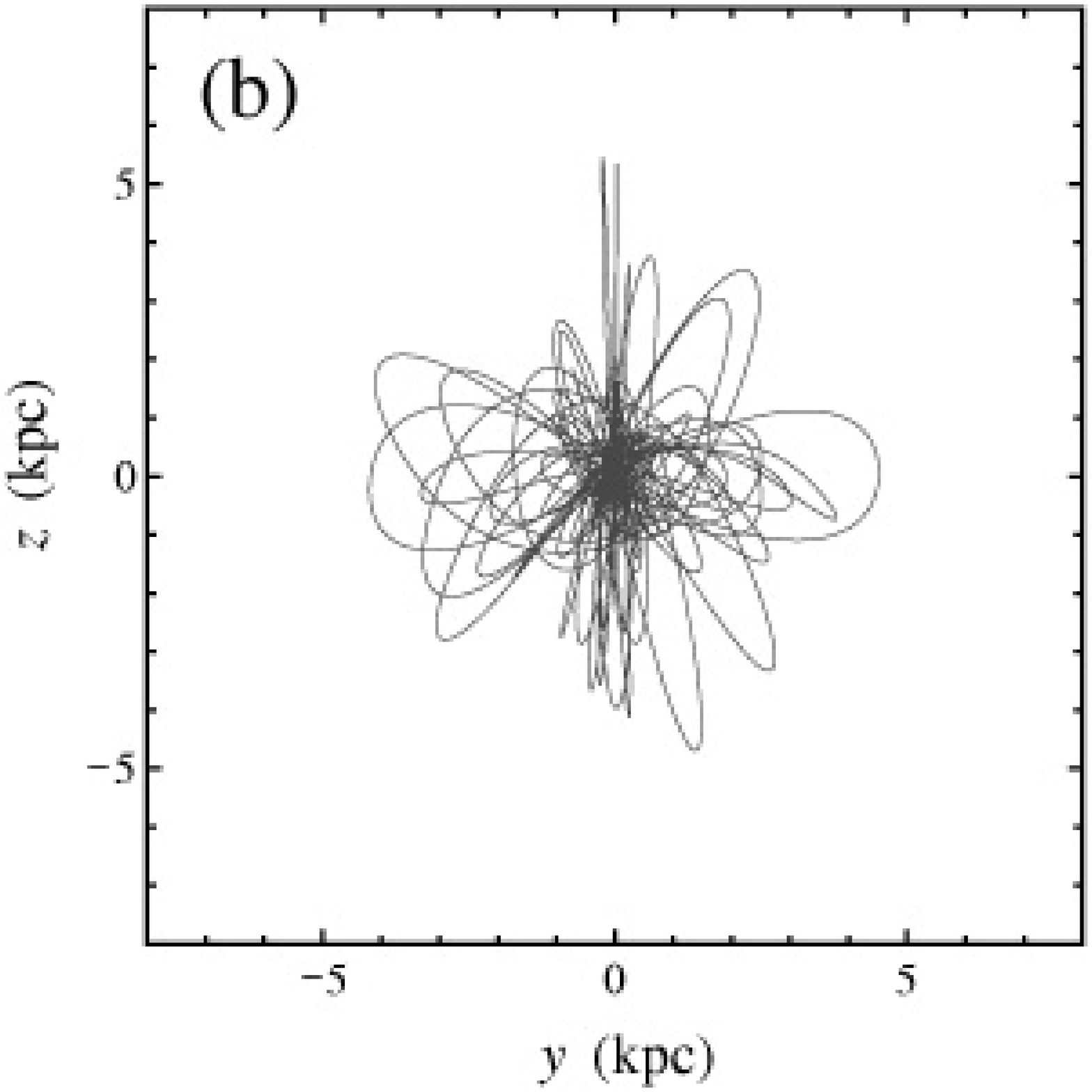} \caption{Same as
 Figure~\ref{fig:orbitA1} but for model~C3. \label{fig:orbitC3}}
\end{figure}

\begin{figure}
\epsscale{1.0} \plottwo{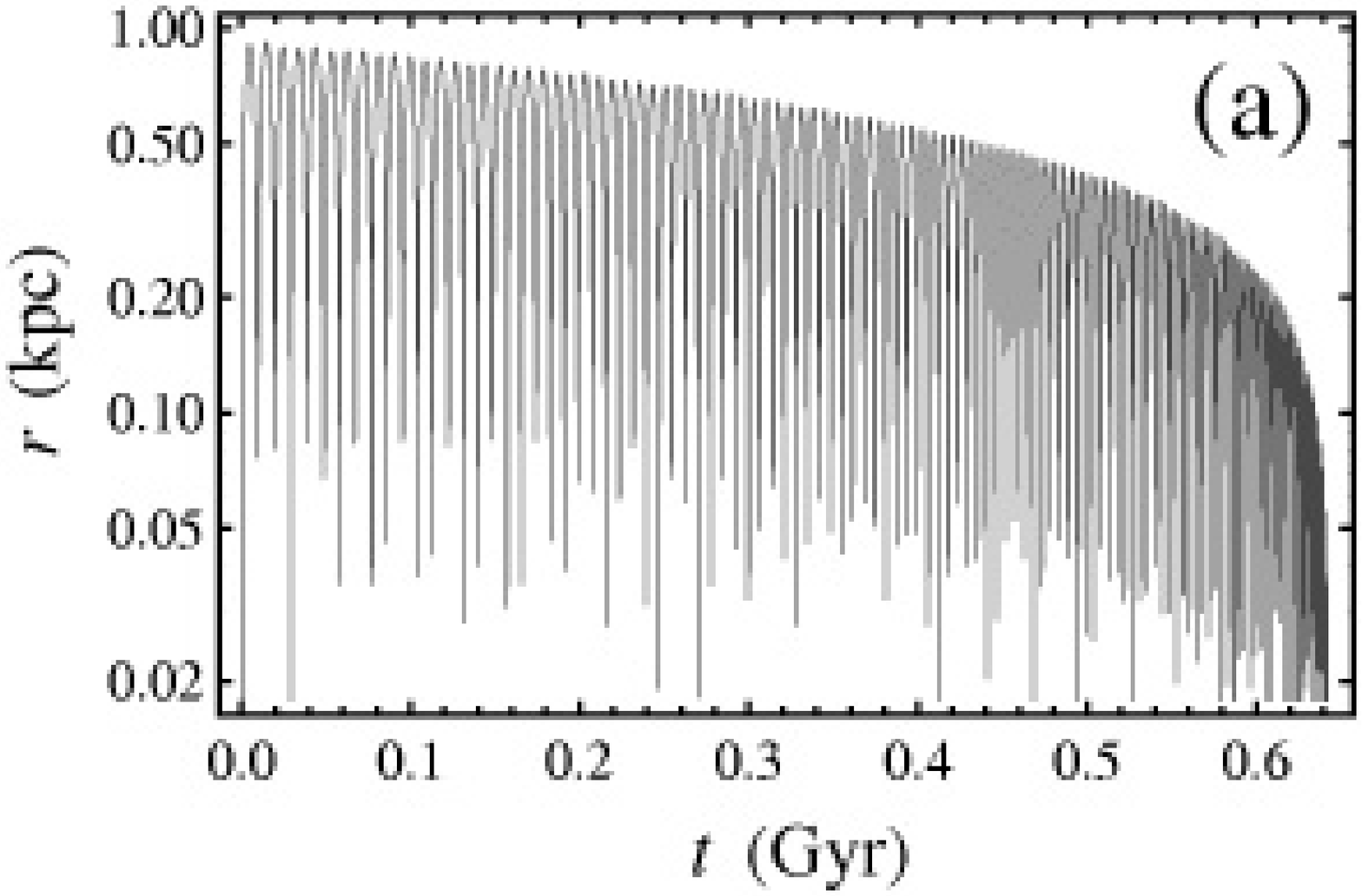}{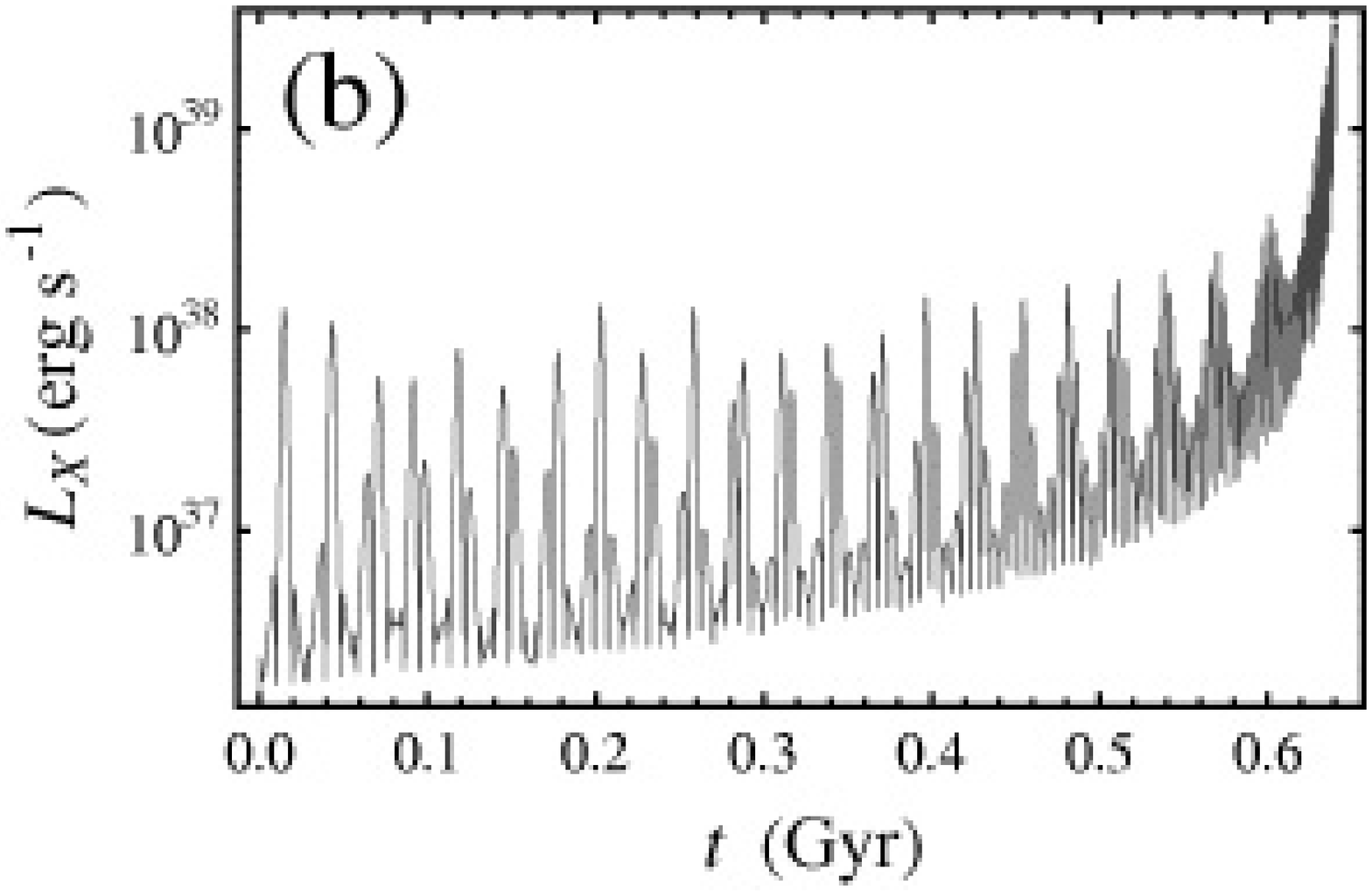} \caption{(a) The distance from
 the galactic center and (b) the luminosity of the SMBH for model~A1 for
 $0<t<t_{\rm df}$ when $\theta=60\arcdeg$. \label{fig:A1}}
\end{figure}

\begin{figure}
\epsscale{1.0} \plottwo{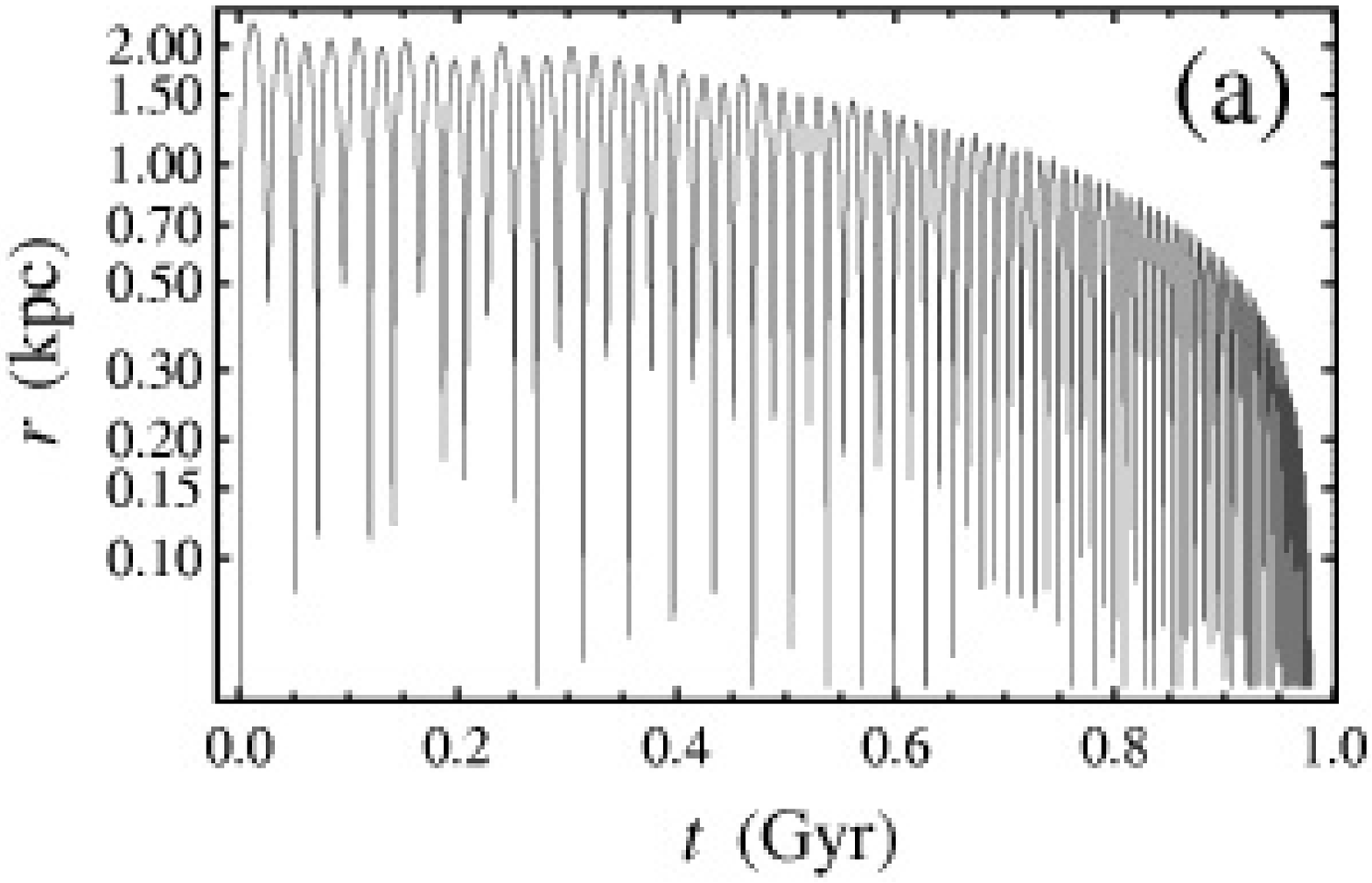}{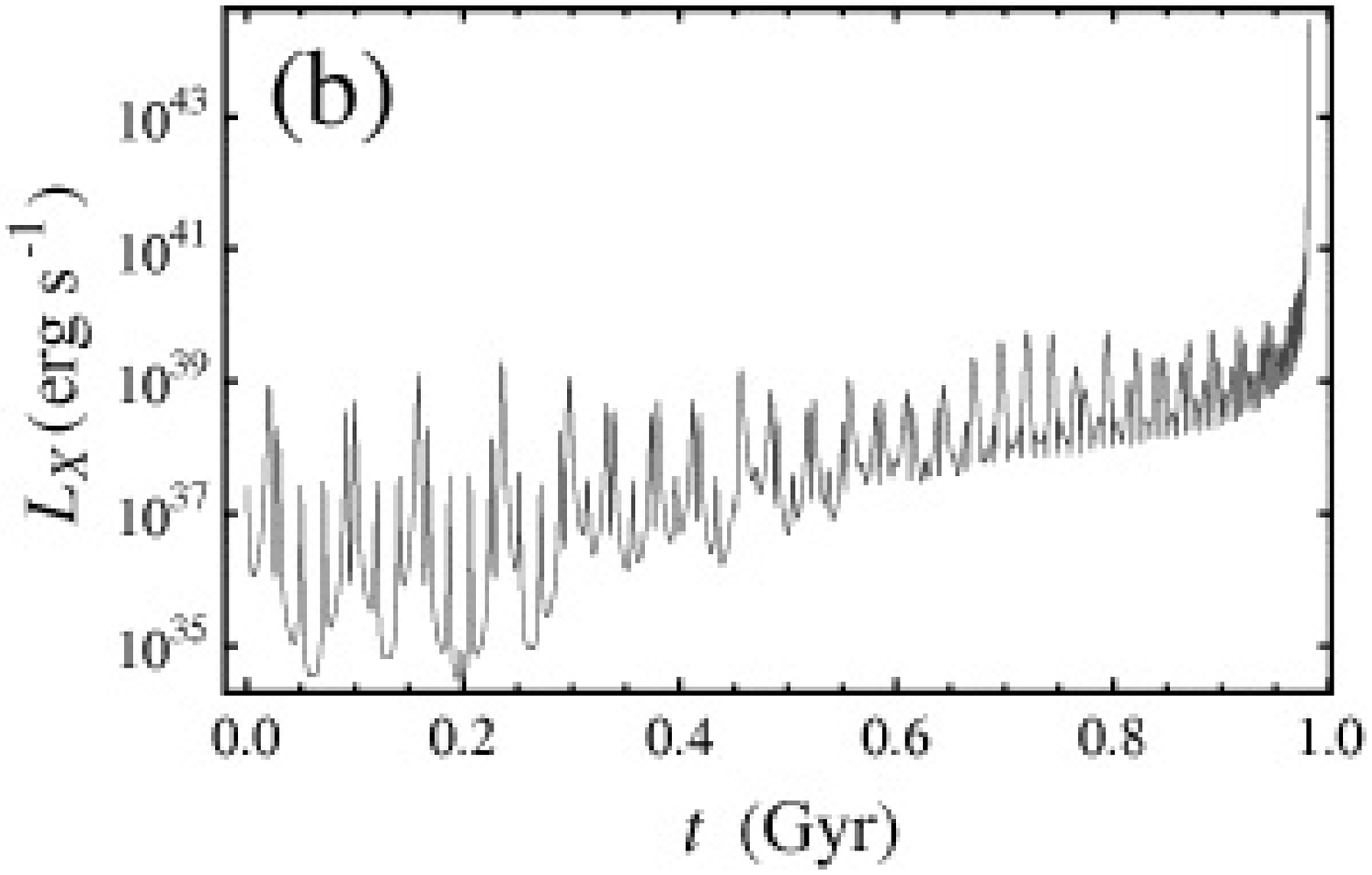} \caption{Same as
 Figure~\ref{fig:A1} but for model~B2. \label{fig:B2}}
\end{figure}

\begin{figure}
\epsscale{1.0} \plottwo{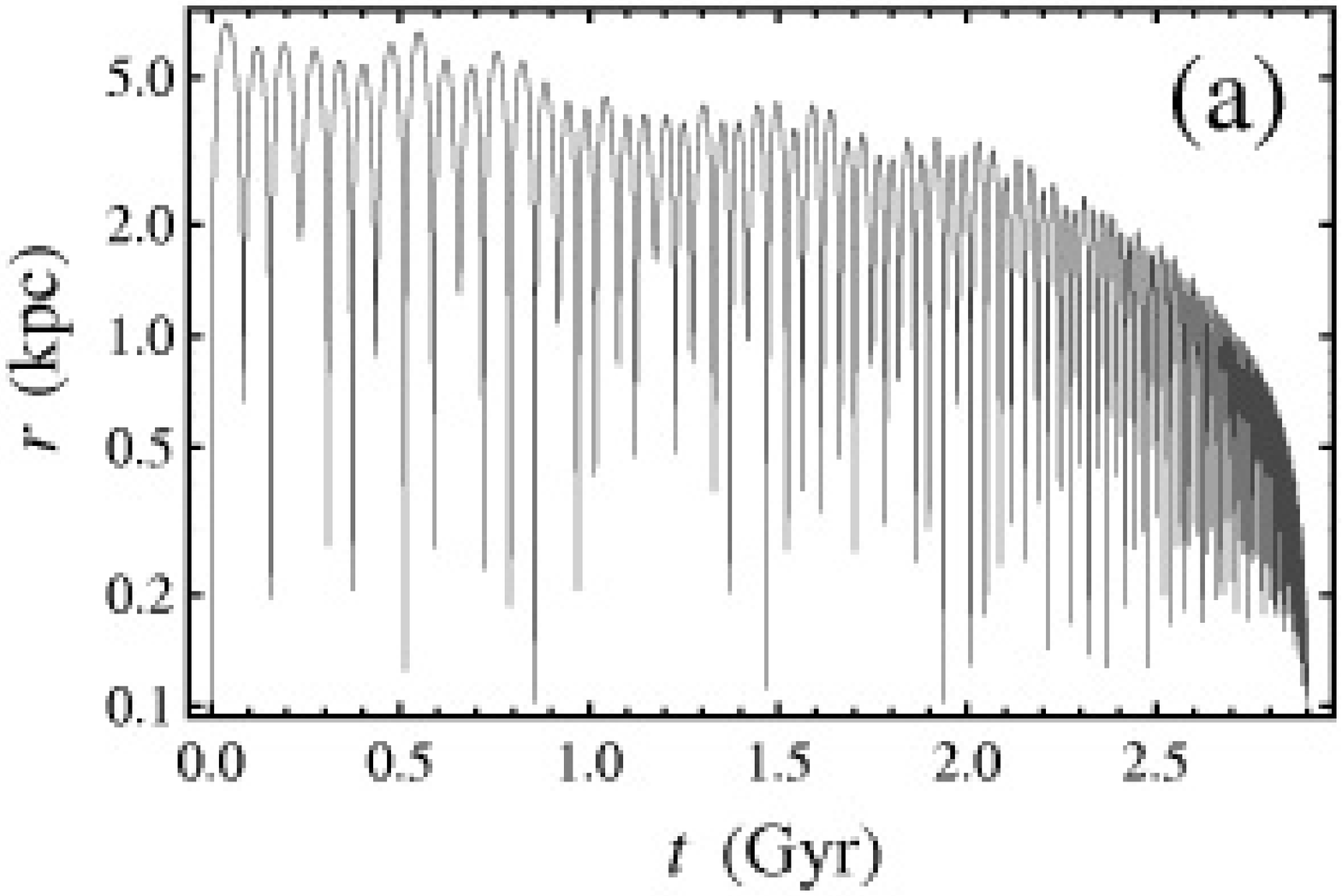}{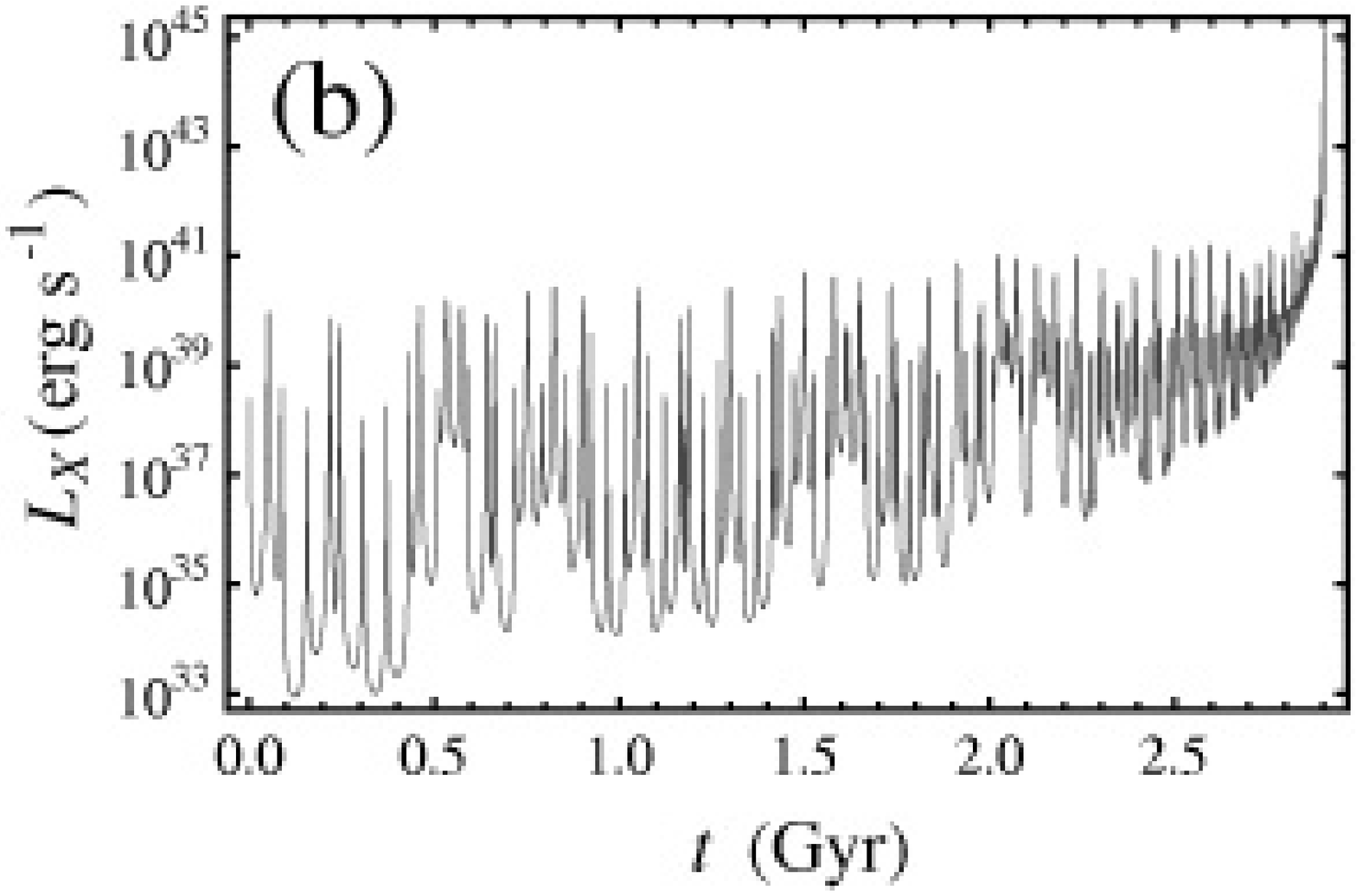} \caption{Same as
 Figure~\ref{fig:A1} but for model~C3.\label{fig:C3}}
\end{figure}

We stop the calculation if (i) $t=10$~Gyr or if (ii) $r<10$~pc and
$v<1\rm\: km\: s^{-1}$ is satisfied. We define the time when the
condition (ii) is satisfied as $t_{\rm
df}$. Figures~\ref{fig:orbitA1}--\ref{fig:orbitC3} show the trajectories
of the SMBHs for models~A1, B2, and C3, respectively. Although the SMBHs
are ejected on the $x$-$z$ plane at $t=0$, they are not confined to the
plane because of the circulation of the galactic disk and dynamical
friction. Figures~\ref{fig:A1}--\ref{fig:C3} show the distance from the
galactic center ($r$) and the luminosity of the SMBHs ($L_X$) for
models~A1, B2, and C3, respectively ($0<t<t_{\rm df}$ and
$\theta=60\arcdeg$). The distance gradually decreases through the
dynamical friction. The infall of the SMBHs accelerates as $t$
approaches $t_{\rm df}$. The luminosity on average increases as the
SMBHs spiral into the galactic center, where $\rho_{\rm ISM}$ is
large. In Table~\ref{tab:par}, we present the maximum distance from the
center of the galaxy when $\theta=60\arcdeg$ ($r_{\rm max,60}$). We note
that the maximum radius is not much dependent on $\theta$.

\begin{figure}
\epsscale{0.5}\plotone{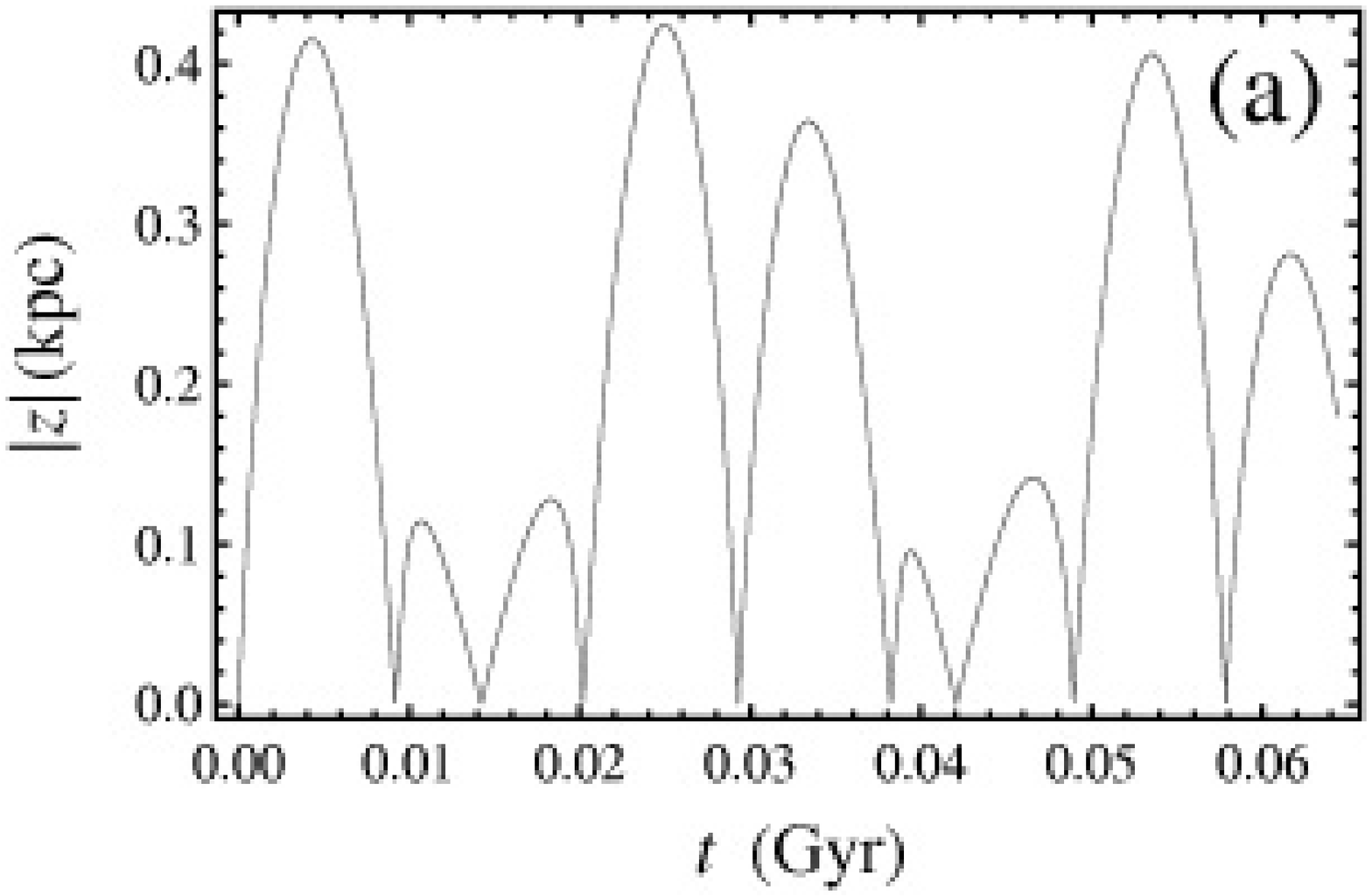}
\vspace{3mm}
\epsscale{0.5}\plotone{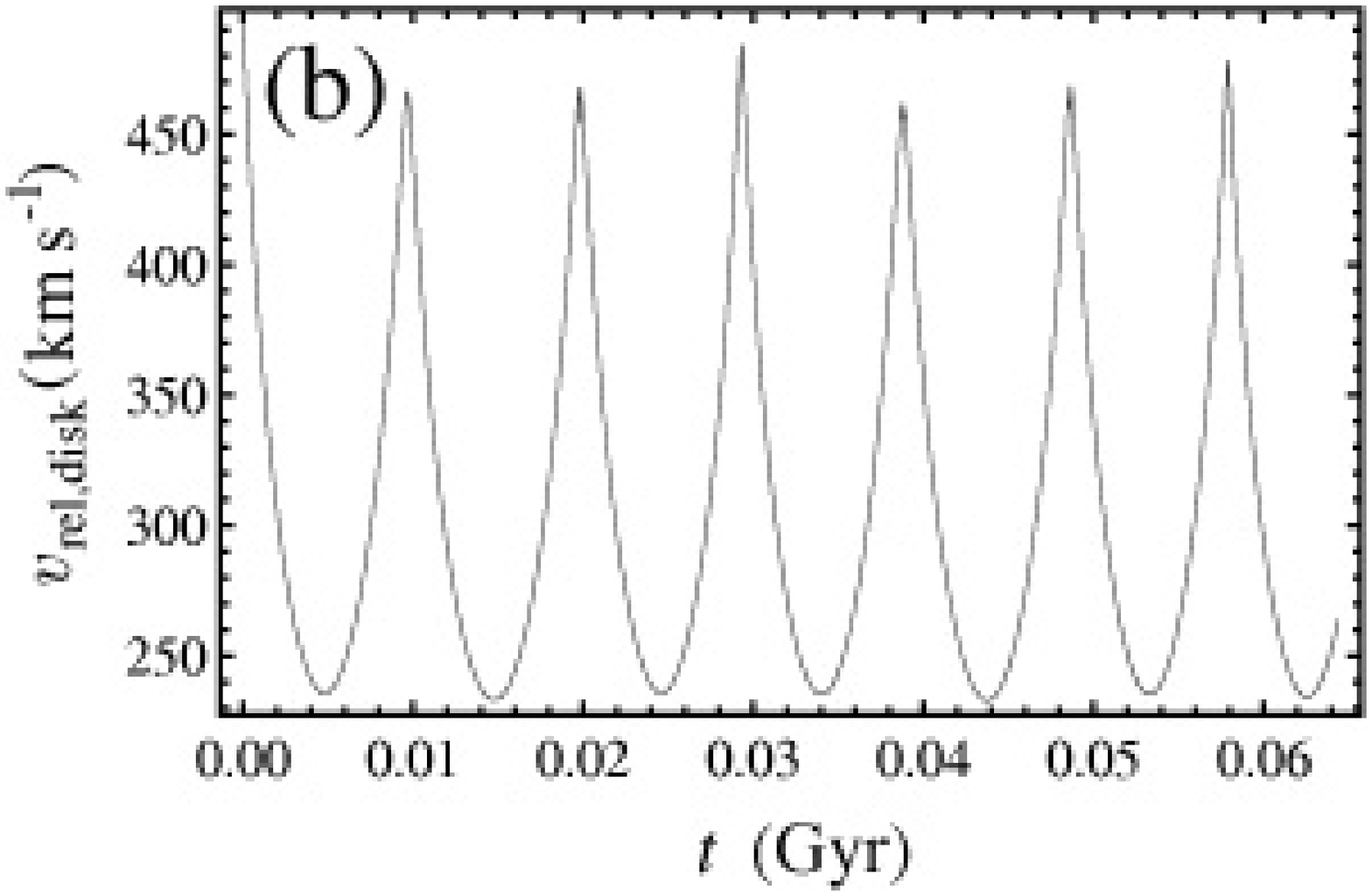} 
\vspace{3mm}
\epsscale{0.5}\plotone{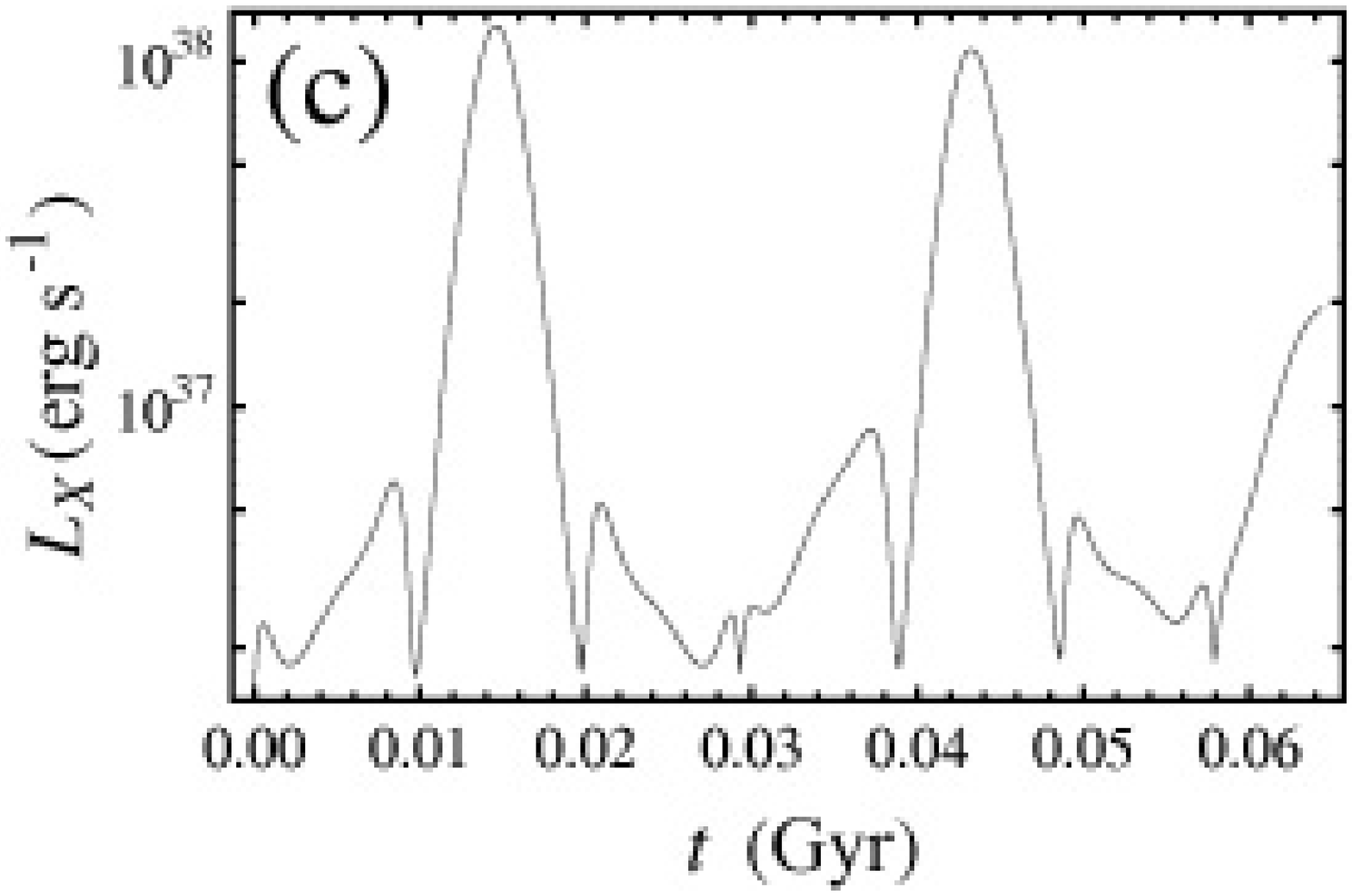} \caption{(a) The distance from the
galactic plane, (b) the relative velocity between the SMBH and the
galactic disk, and (c) the luminosity of the SMBH for model~A1 for
$0<t<0.1\: t_{\rm df}$ when $\theta=60\arcdeg$. \label{fig:A1s}}
\end{figure}

\begin{figure}
\epsscale{0.5}\plotone{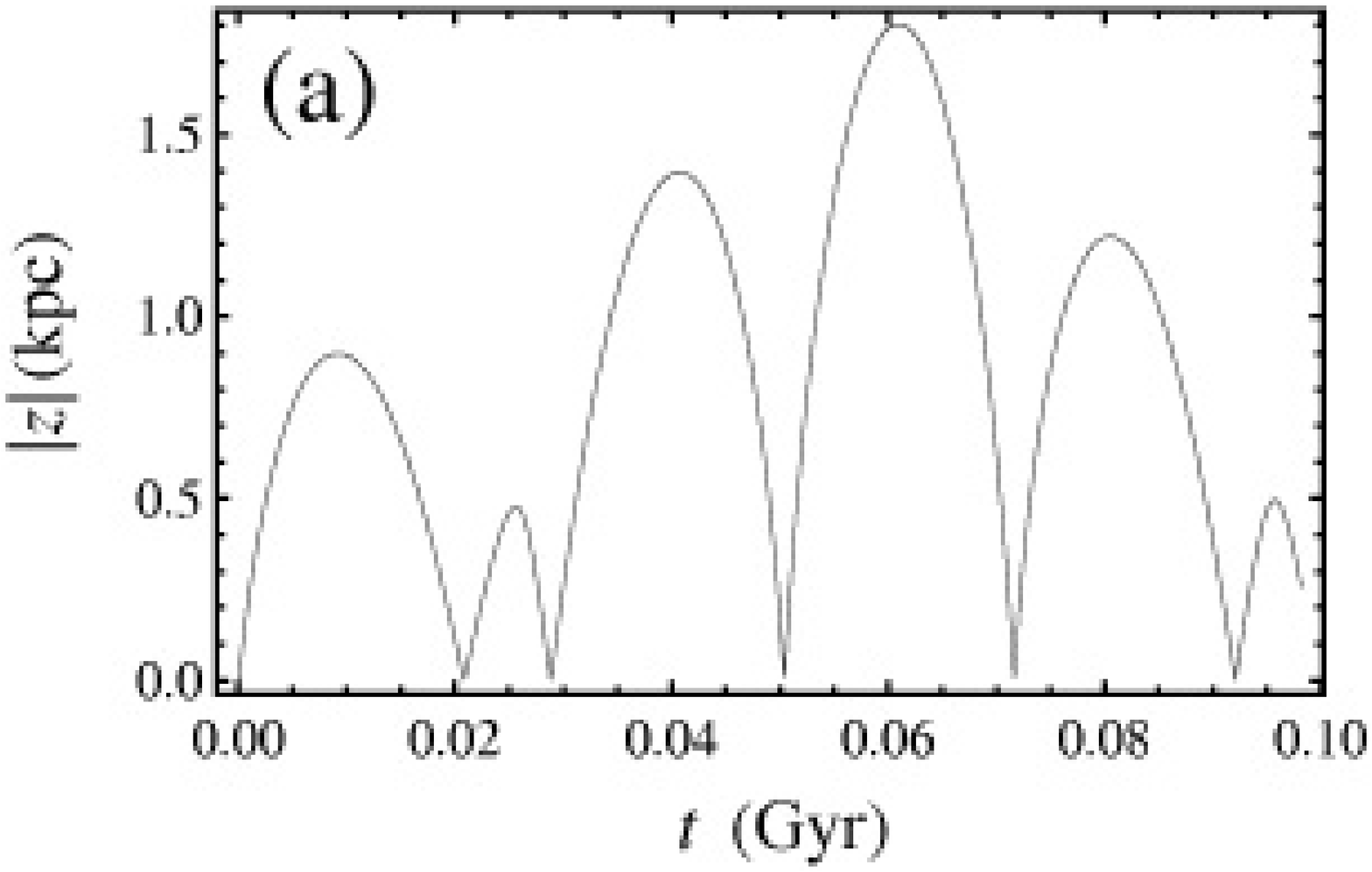}
\vspace{3mm}
\epsscale{0.5}\plotone{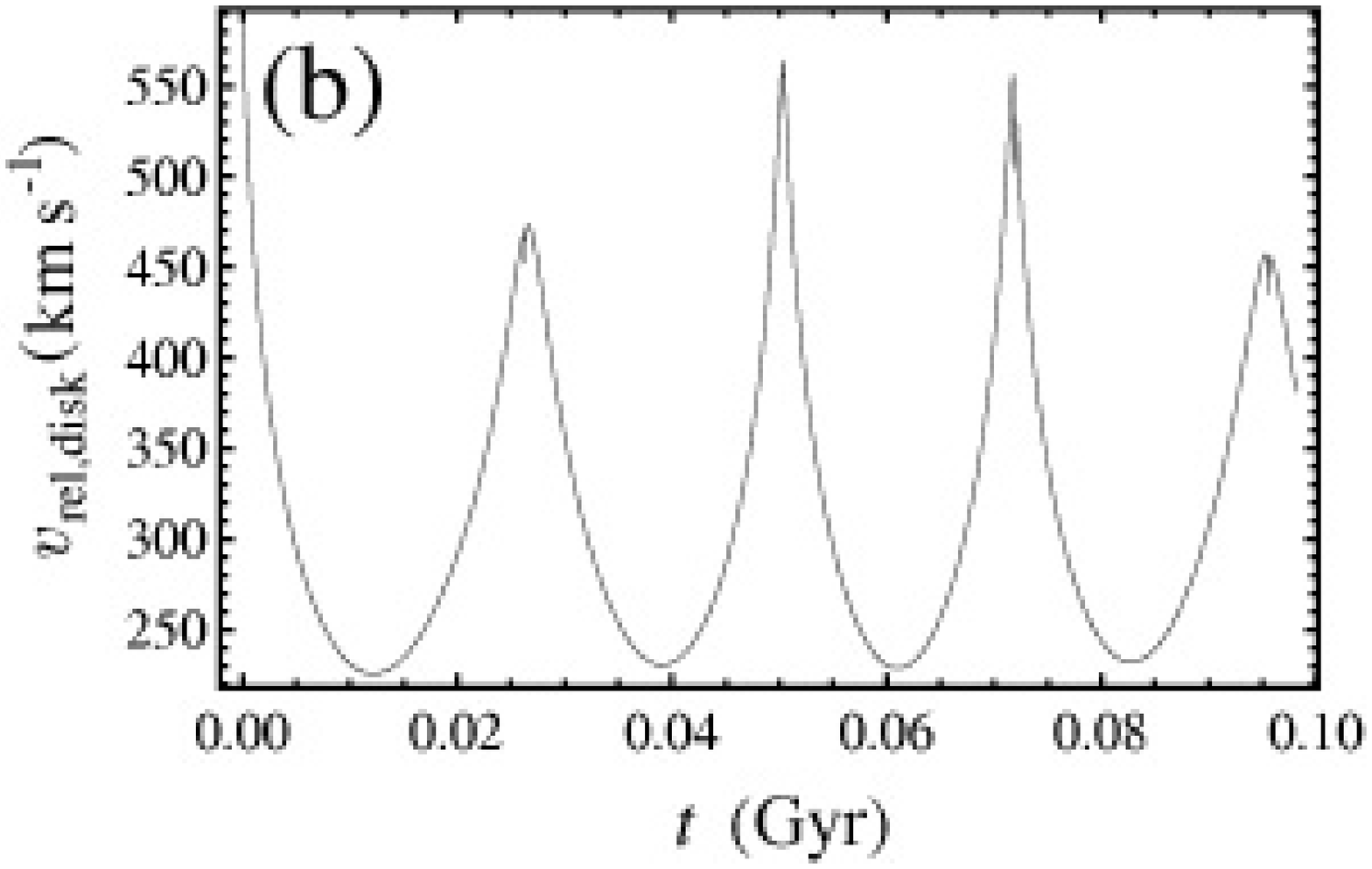} 
\vspace{3mm}
\epsscale{0.5}\plotone{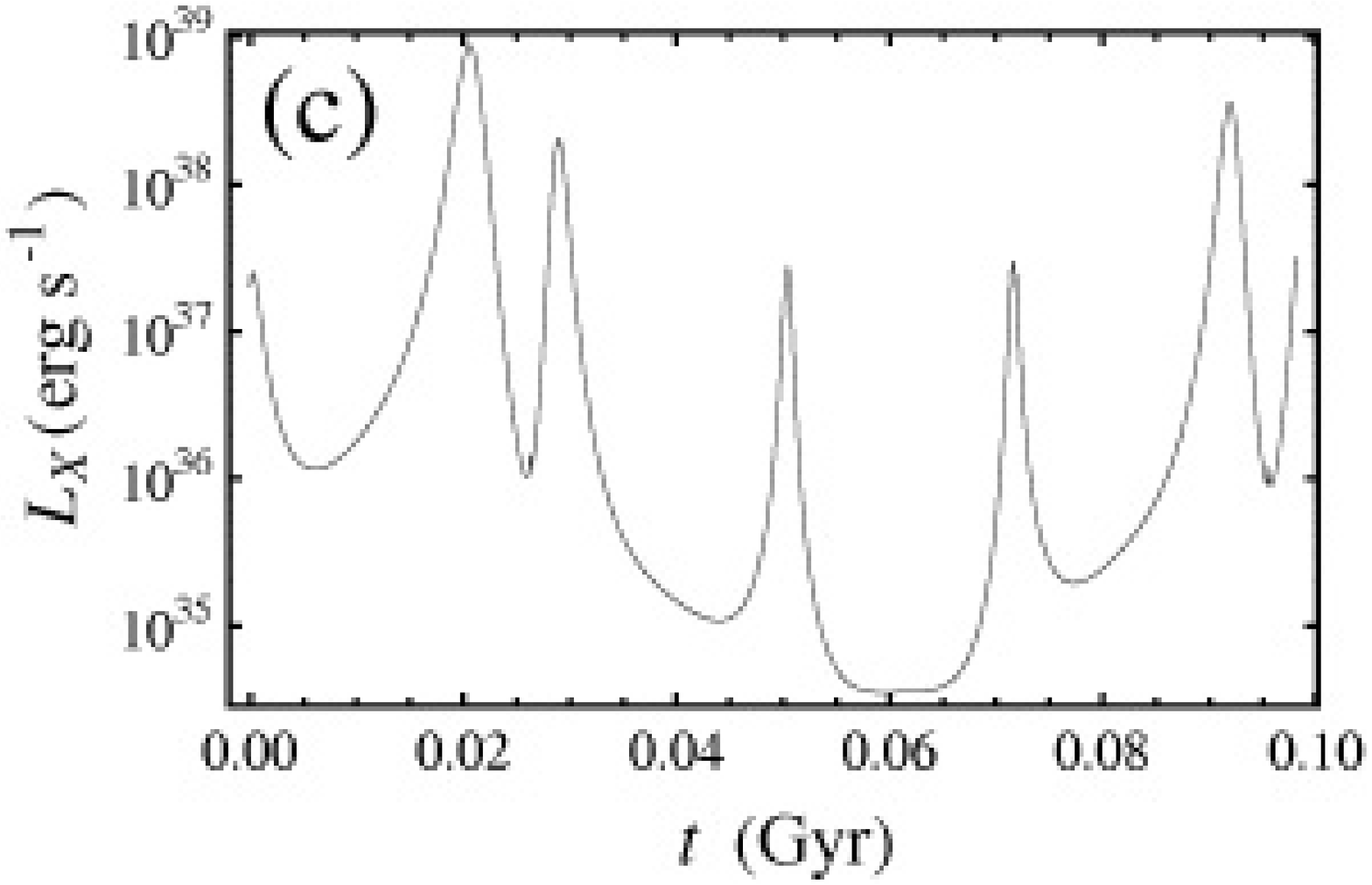}
\caption{Same as
 Figure~\ref{fig:A1s} but for model~B2. \label{fig:B2s}}
\end{figure}

\begin{figure}
\epsscale{0.5}\plotone{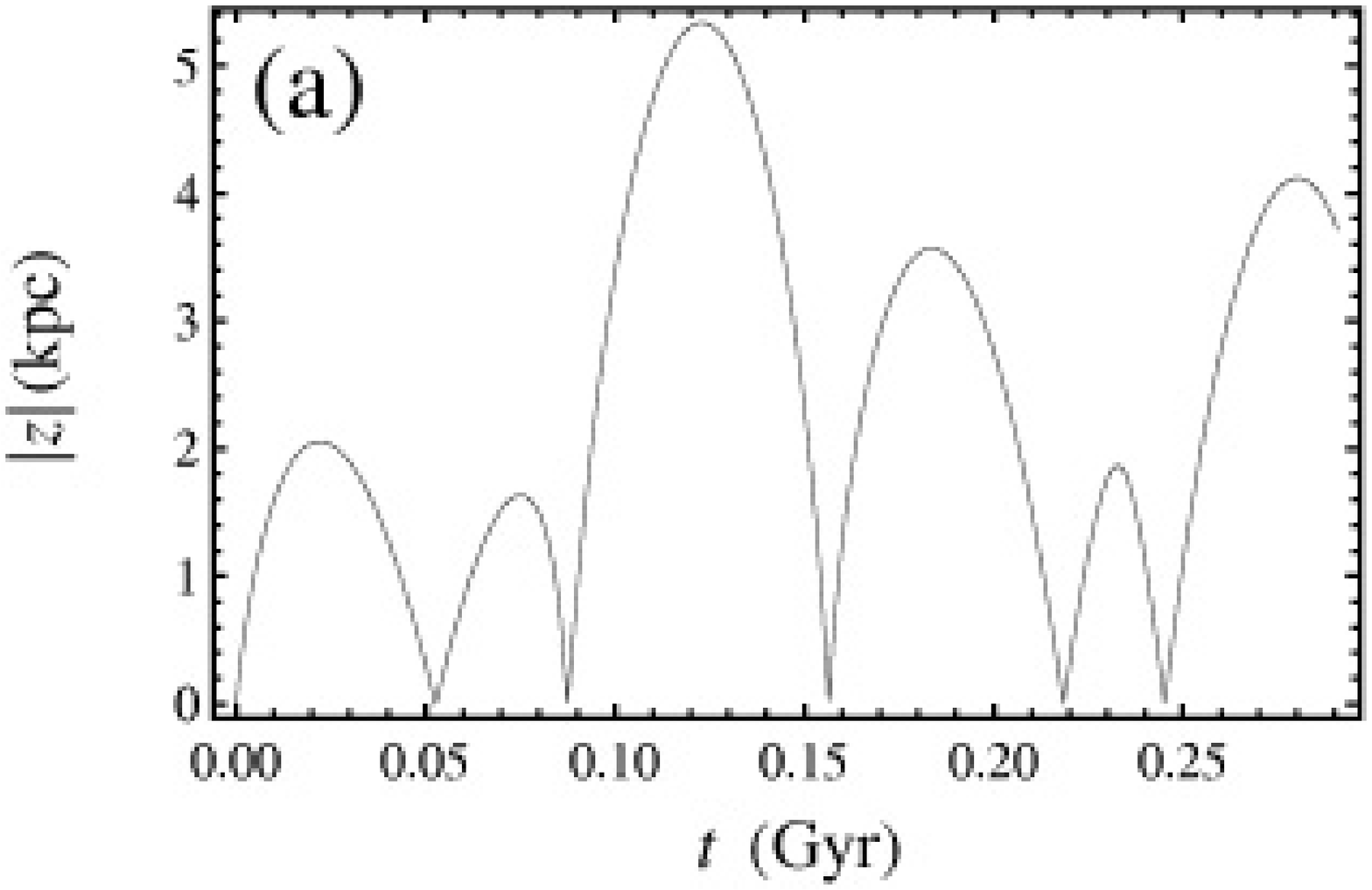}
\vspace{3mm}
\epsscale{0.5}\plotone{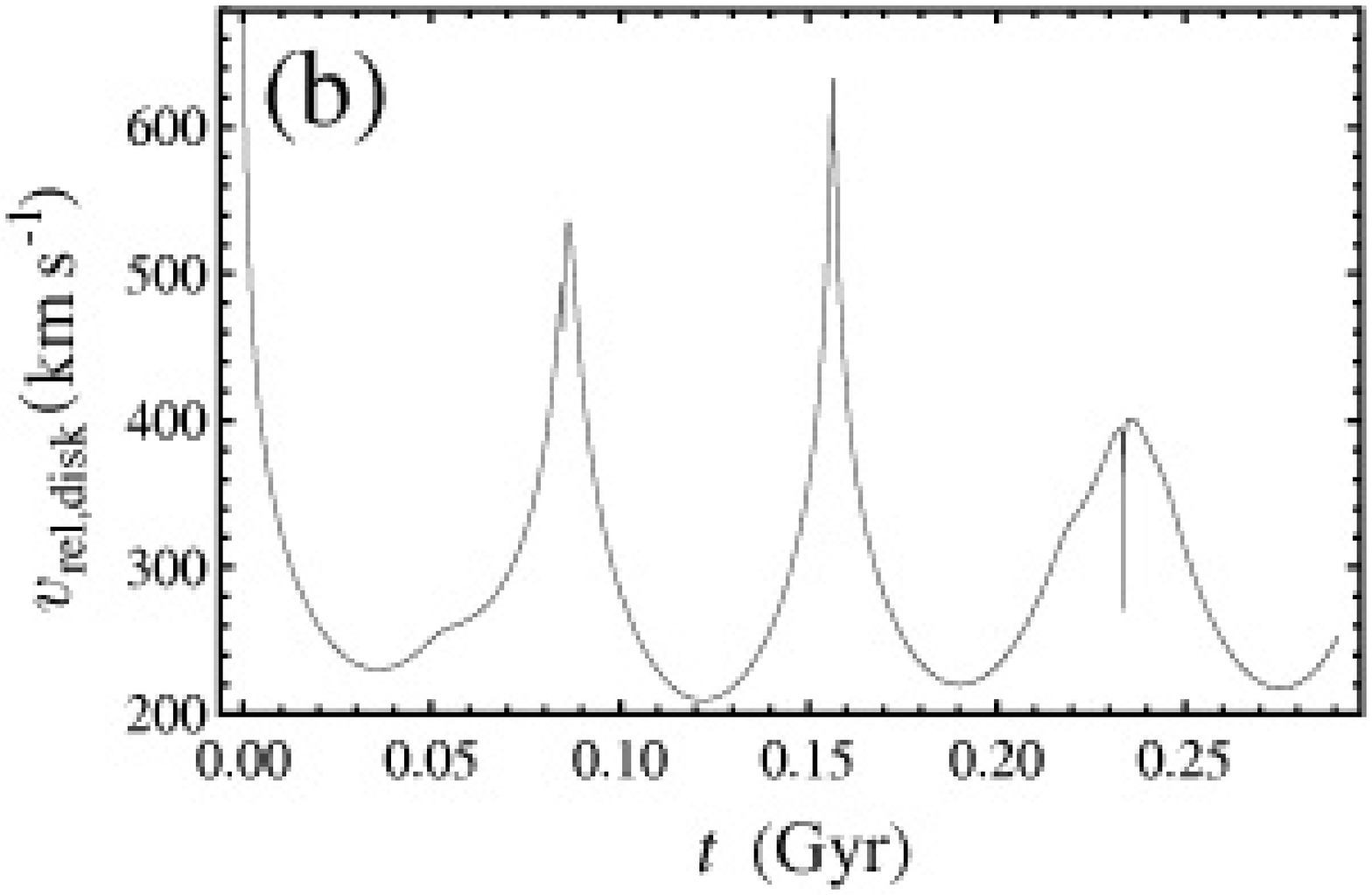} 
\vspace{3mm}
\epsscale{0.5}\plotone{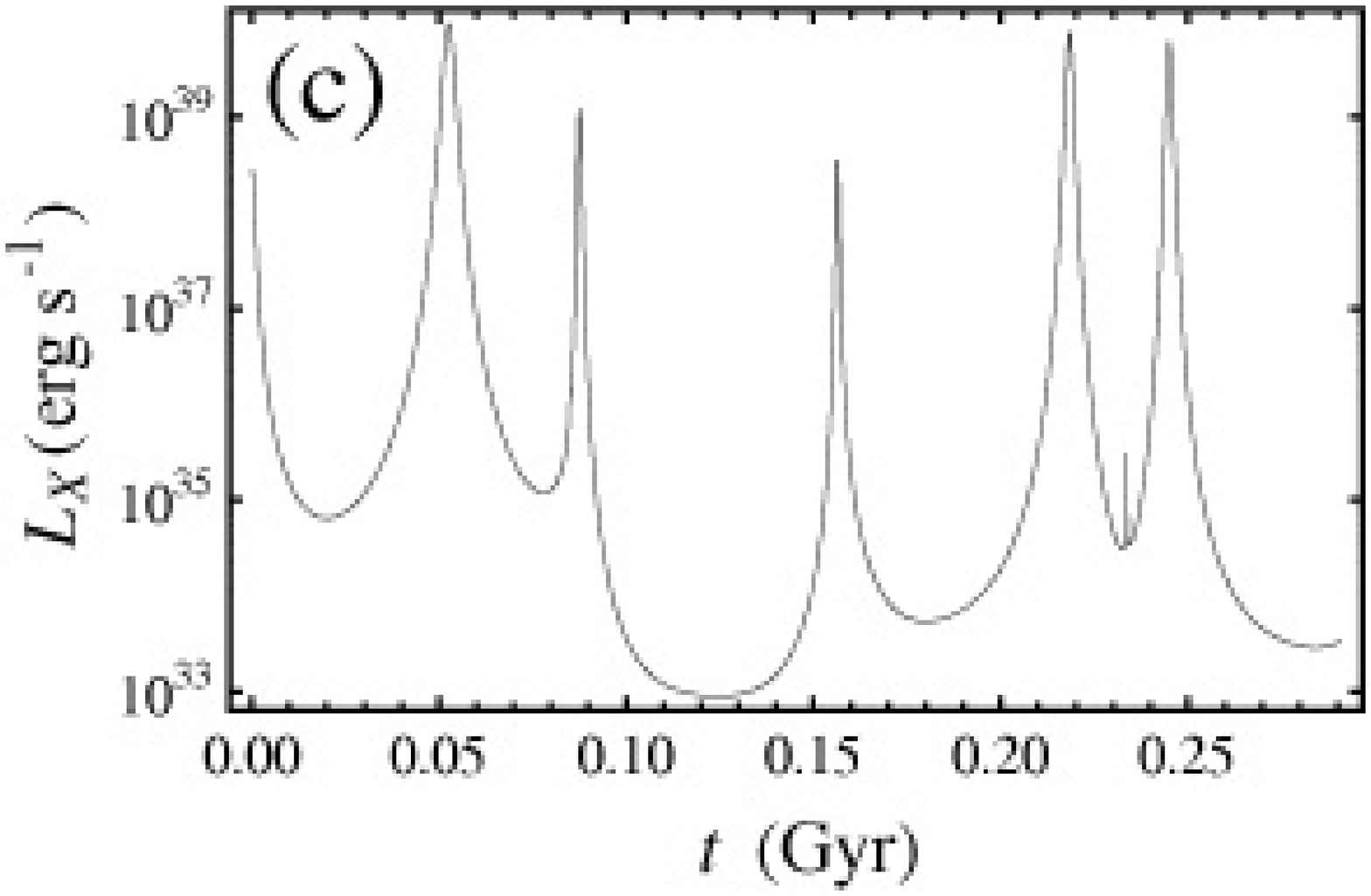}
\caption{Same as
 Figure~\ref{fig:A1s} but for model~C3. \label{fig:C3s}}
\end{figure}

Figures~\ref{fig:A1s}--\ref{fig:C3s} show the evolutions of $|z|$,
$v_{\rm rel,disk}$, and $L_X$ for $0<t<0.1\: t_{\rm df}$ for models~A1,
B2, and C3, respectively ($\theta=60\arcdeg$). The luminosity of the
SMBHs ($L_X$) increases instantaneously, when they pass the galactic
disk. The heights of the spikes in Figure~\ref{fig:A1s}b,
\ref{fig:B2s}b, and \ref{fig:C3s}b are uneven. This is because the
luminosity $L_X$ depends on both $v_{\rm rel,disk}$ and $\rho_{\rm ISM}$
(see equation~[\ref{eq:dotm}]), and the latter strongly depends on
$z$. In Table~\ref{tab:par}, we present the maximum X-ray luminosity of
the SMBHs for $\theta=60\arcdeg$ and $t<0.2\: t_{\rm df}$ ($L_{\rm
max,60}$). The luminosity $L_{\rm max,60}$ is larger for larger $m_{\rm
BH}$ and smaller $v_0$. In some of the models of $m_{\rm BH}\geq 1\times
10^7\: M_\sun$, $L_{\rm max,60}$ reaches $3\times 10^{39}\rm\: erg\:
s^{-1}$, which is the Eddington luminosity of a stellar mass black hole
($\sim 20\: M_\sun$) and is often used as a threshold of ULXs
\citep{col99,mak00,mus04}. When $m_{\rm BH}$ and $v_0$ are given and
$\theta$ is not fixed, the X-ray luminosity before the SMBHs are
affected by dynamical friction tends to be larger when $\theta$ is
closer to $90\arcdeg$, because their trajectories are included in the
galactic disk, where $\rho_{\rm ISM}$ is large. However, the tendency is
not clear when $\theta\la 80\arcdeg$, because their orbits are scattered
in the asymmetric potential of the galaxy. Thus, the maximum luminosity
does not much depend on $\theta$. Figures~\ref{fig:A1s}--\ref{fig:C3s}
indicate that when the SMBH is especially bright, the relative velocity
between the SMBH and the surrounding ISM (or stars) is $v_{\rm rel,
disk}\sim v_{\rm cir}(\sim 220\:\rm km\: s^{-1})$, which means that the
SMBH passes the apocenter of its orbit ($v\sim 0$) close to the galactic
plane. It could be used as a clue to find the traveling SMBH
observationally, if atomic line emission associated with the X-ray
source is detected and the velocity is estimated through the Doppler
shift.

In Table~\ref{tab:par}, we present the average of $t_{\rm df}$, which is
referred to as $\langle t_{\rm df}\rangle$; we calculate 30 orbits and
corresponding $t_{\rm df}$ by changing $\theta$ from $3\arcdeg$ to
$90\arcdeg$ by $3\arcdeg$ at a time, and average $t_{\rm df}$ by
$\theta$, weighting with $\sin\theta$. Table~\ref{tab:par} shows that
$\langle t_{\rm df}\rangle\ga 0.1$~Gyr for models A1--C4, and that
$\langle t_{\rm df}\rangle$ is smaller for larger $m_{\rm BH}$ and
smaller $v_0$.

Following Paper~I, we estimate the probability of observing SMBHs with
luminosities larger than a threshold luminosity $L_{\rm th}$, assuming
that SMBHs are ejected in random directions at the centers of
galaxies. For each model, we calculate 30 evolutions of the luminosity
by changing $\theta$ from $3\arcdeg$ to $90\arcdeg$ by $3\arcdeg$ at a
time. Then, we obtain the period during which the relation $L_{\rm
X}>L_{\rm th}$ is satisfied for each $\theta$, and divide the period by
$t_{\rm df}$. This is the fraction of the period during which the black
hole luminosity becomes larger than $L_{\rm th}$. We refer to this
fraction as $f(\theta)$. We average $f(\theta)$ by $\theta$, weighting
with $\sin\theta$, and obtain the probability of observing SMBHs with
$L_{\rm X}>L_{\rm th}$. In Table~\ref{tab:par}, we present the
probability $P_{3e39}$ when $L_{\rm th}=3\times 10^{39}\rm\: erg\:
s^{-1}$. For models A1--C4, $P_{3e39}=0.0018$--0.58.

We also estimate the age-corrected probability of observing SMBHs with
$L_{\rm X}>L_{\rm th}=3\times 10^{39}\rm\: erg\: s^{-1}$, which is
obtained by averaging $\min[t_{\rm df},t_{\rm age}] f(\theta)/t_{\rm
age}$ by $\theta$, weighting with $\sin\theta$, where $t_{\rm age}$ is
the age of a galaxy and we assume that $t_{\rm age}=10$~Gyr. We refer to
the age-corrected probability as $\tilde{P}_{\rm 3e39}$ and show it in
Table~\ref{tab:par}.

\section{Discussion}
\label{sec:discussion}

We have found that a SMBH that had been ejected from the center of a
disk galaxy could be observed in the galactic disk with an X-ray
luminosity of $L_{\rm X}\ga 3\times 10^{39}\rm\: erg\: s^{-1}$. The
luminosity gradually increases as the SMBH settles down to the galactic
center through dynamical friction.

In \S~\ref{sec:results}, we follow the evolution until the SMBH spirals
down to $r=10$~pc. However, if $r$ is too small, the SMBH cannot be
discriminated from the one that would have been sitting at the galactic
center without being affected by a recoil. Therefore, we estimate the
probability of observing SMBHs with $L_X>L_{\rm th}=3\times 10^{39}\rm\:
erg\: s^{-1}$ and $r>1$~kpc, and call it $P_{\rm 1,3e39}$. We also
calculate the time age-corrected one ($\tilde{P}_{\rm 1,3e39}$). Since
$P_{\rm 1,3e39}$ and $\tilde{P}_{\rm 1,3e39}$ are derived by adding
another condition $r>1$~kpc to $P_{\rm 3e39}$ and $\tilde{P}_{\rm
3e39}$, respectively, it is natural that $P_{\rm 1,3e39}\leq P_{\rm
3e39}$ and $\tilde{P}_{\rm 1,3e39}\leq\tilde{P}_{\rm 3e39}$
(Table~\ref{tab:par}).

\begin{figure}
\epsscale{0.5}\plotone{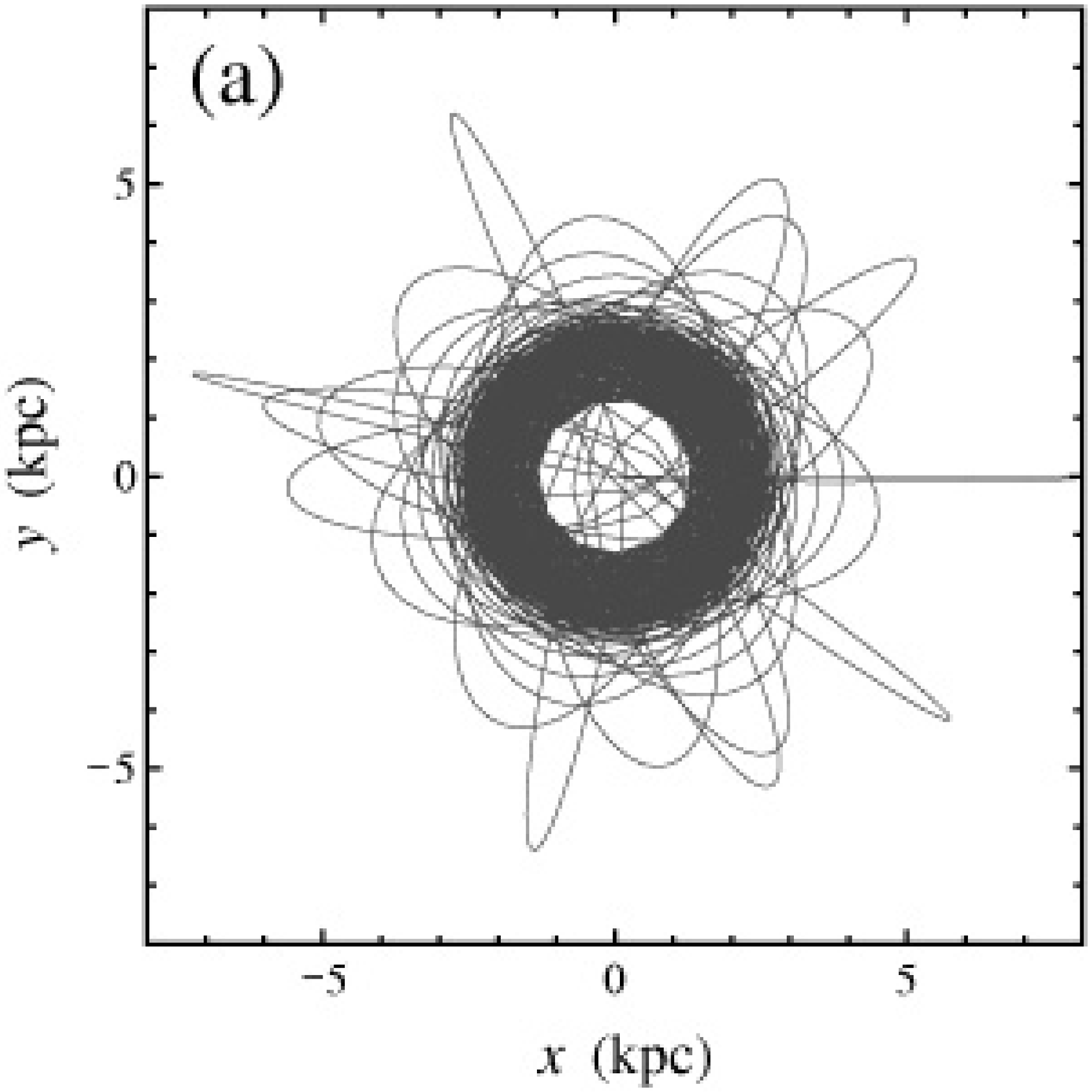}
\vspace{3mm}
\epsscale{0.5}\plotone{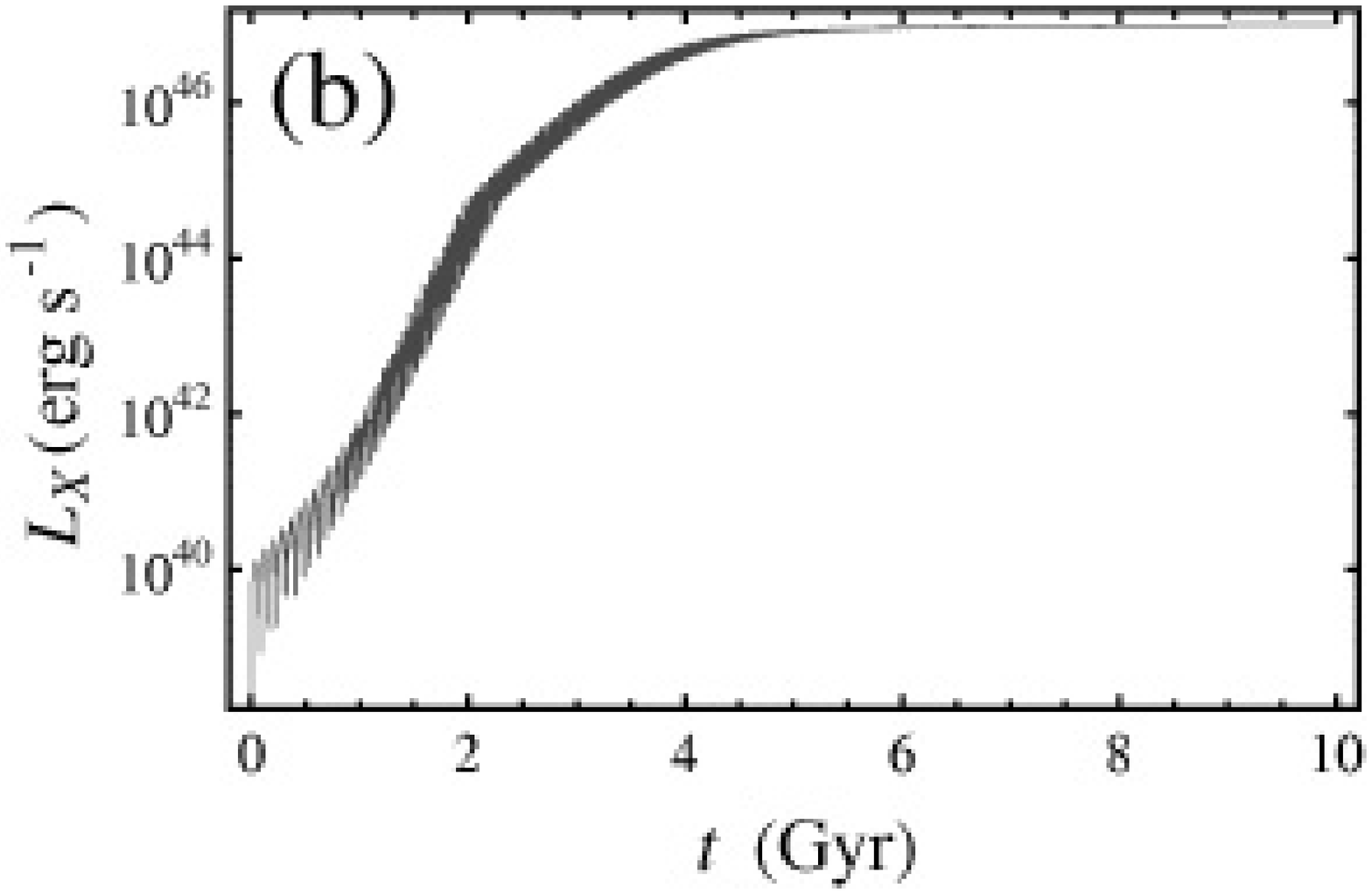} \caption{(a) The trajectory in the
galactic disk plane ($z=0$) and (b) the luminosity of the SMBH for
model~C3 for $0<t<10$~Gyr when $\theta=90\arcdeg$. \label{fig:C3_90}}
\end{figure}

As is mentioned in \S~\ref{sec:models}, dynamical friction of a massive
point particle orbiting in a disk galaxy has not been studied very
much. Thus, there is some uncertainty about the Coulomb logarithm we
should take. Therefore, we change the value of $\ln\Lambda_{\rm disk}$
to estimate the uncertainty. Models C$'$2 and C$'$3 are respectively the
same as models C2 and C3 except for $\ln\Lambda_{\rm disk}$. For these
models, we set $\ln\Lambda_{\rm disk}=1.5$. Table~\ref{tab:par} shows
that there is not much difference between the results of models~C2 and
those of C$'$2. This is because the maximum distances to the apocenters
are $\la 1$~kpc, where the spheroidal component is dominant, and the
SMBH is not much affected by the dynamical friction from the galactic
disk. On the other hand, $P_{\rm 3e39}$, $\tilde{P}_{\rm 3e39}$, $P_{\rm
1,3e39}$, and $\tilde{P}_{\rm 1,3e39}$ for models~C3 and C$'$3 are
significantly different, because the SMBH is ejected outside the
spheroid. The differences are especially made by that of the orbits of
$\theta\sim 90\arcdeg$. When $\theta\sim 90\arcdeg$, the SMBH is ejected
in the galactic disk. If the dynamical friction from the disk is very
effective, the SMBH moves along with the disk ($v_{\rm rel,disk}\sim 0$)
and does not easily fall into the galactic center. This actually happens
for model~C3 ($t_{\rm df}>10$~Gyr when $\theta=90\arcdeg$;
Figure~\ref{fig:C3_90}a). In this case, the SMBH continues to accrete
the ISM in the disk and is bright for a long time
(Figure~\ref{fig:C3_90}b). Since this SMBH is very bright ($L_X\ga
10^{46}\rm\: erg\: s^{-1}$), it could be easily observed if such SMBHs
actually exist. For model~C$'$3, the dynamical friction from the disk is
not strong enough to hold back the SMBH from the infall even when
$\theta=90\arcdeg$.

We also consider the uncertainty of the ISM fraction $f_{\rm
ISM}$. Model~C$''$3 is the same as model~C3 but for $f_{\rm
ISM}=0.1$. For parameters we adopted, the accretion efficiency is
$\dot{m}<0.1\: \dot{m}_{\rm Edd}$ in most cases. Therefore, we obtain
$L_X\propto \eta\dot{m}\propto \dot{m}^2\propto \rho_{\rm ISM}^2\propto
f_{\rm ISM}^2$ (equations~[\ref{eq:dotm}] and~[\ref{eq:Lx}]), which
means that the X-ray luminosity in model~C$''$3 is one fourth of that in
model~C3. Accordingly, $P_{\rm 3e39}$, $\tilde{P}_{\rm 3e39}$, $P_{\rm
1,3e39}$, and $\tilde{P}_{\rm 1,3e39}$ in model~C$''$3 are smaller than
those in model~C3, respectively. However, the differences are not large,
because $L_X$ changes rapidly.

\begin{figure}
\epsscale{0.5}\plotone{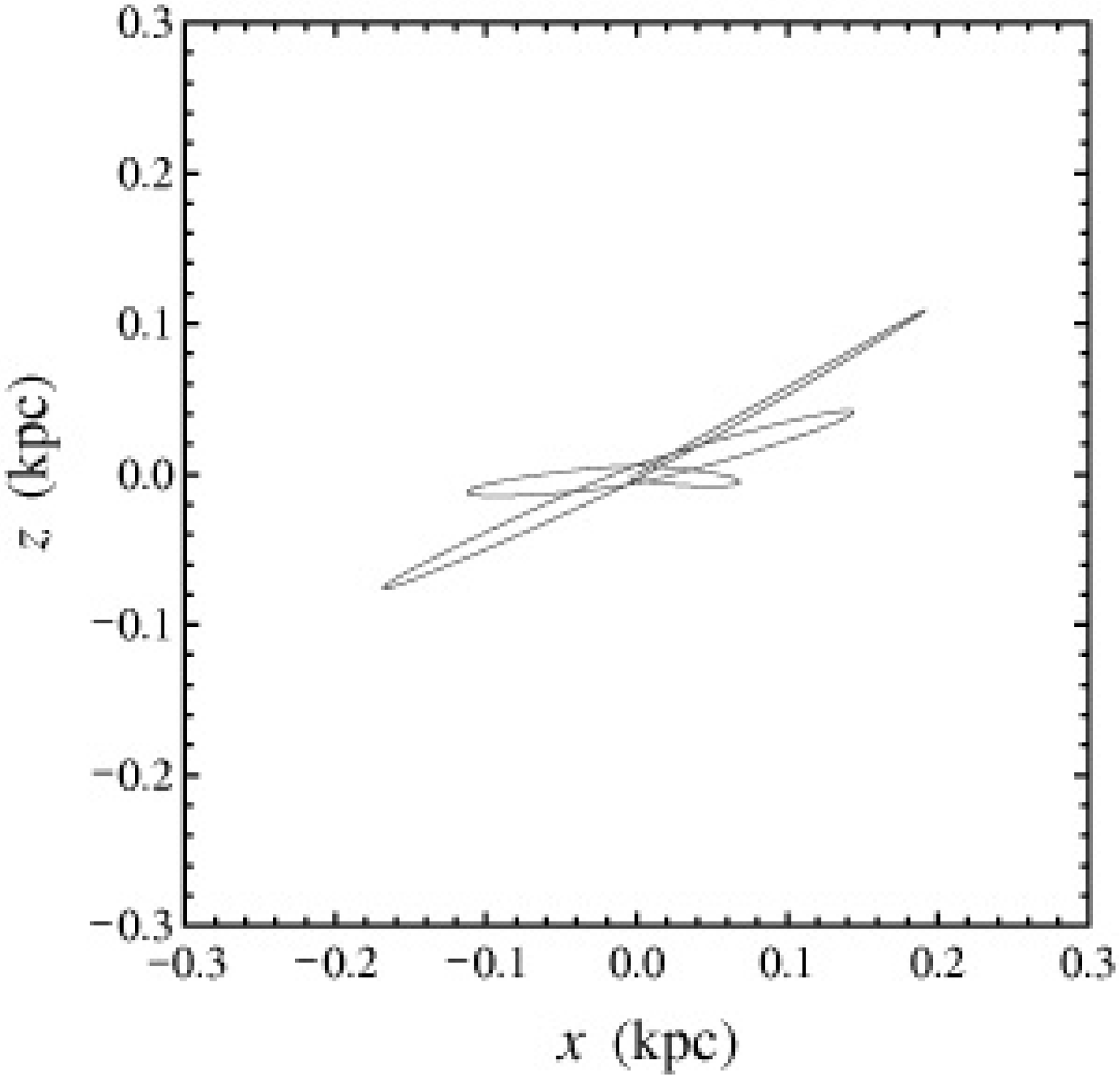}
\caption{The trajectory of the
SMBH for model~b0 for $0<t<t_{\rm df}$ when
$\theta=60\arcdeg$. \label{fig:orbitb0}}
\end{figure}

\begin{figure}
\epsscale{0.5}\plotone{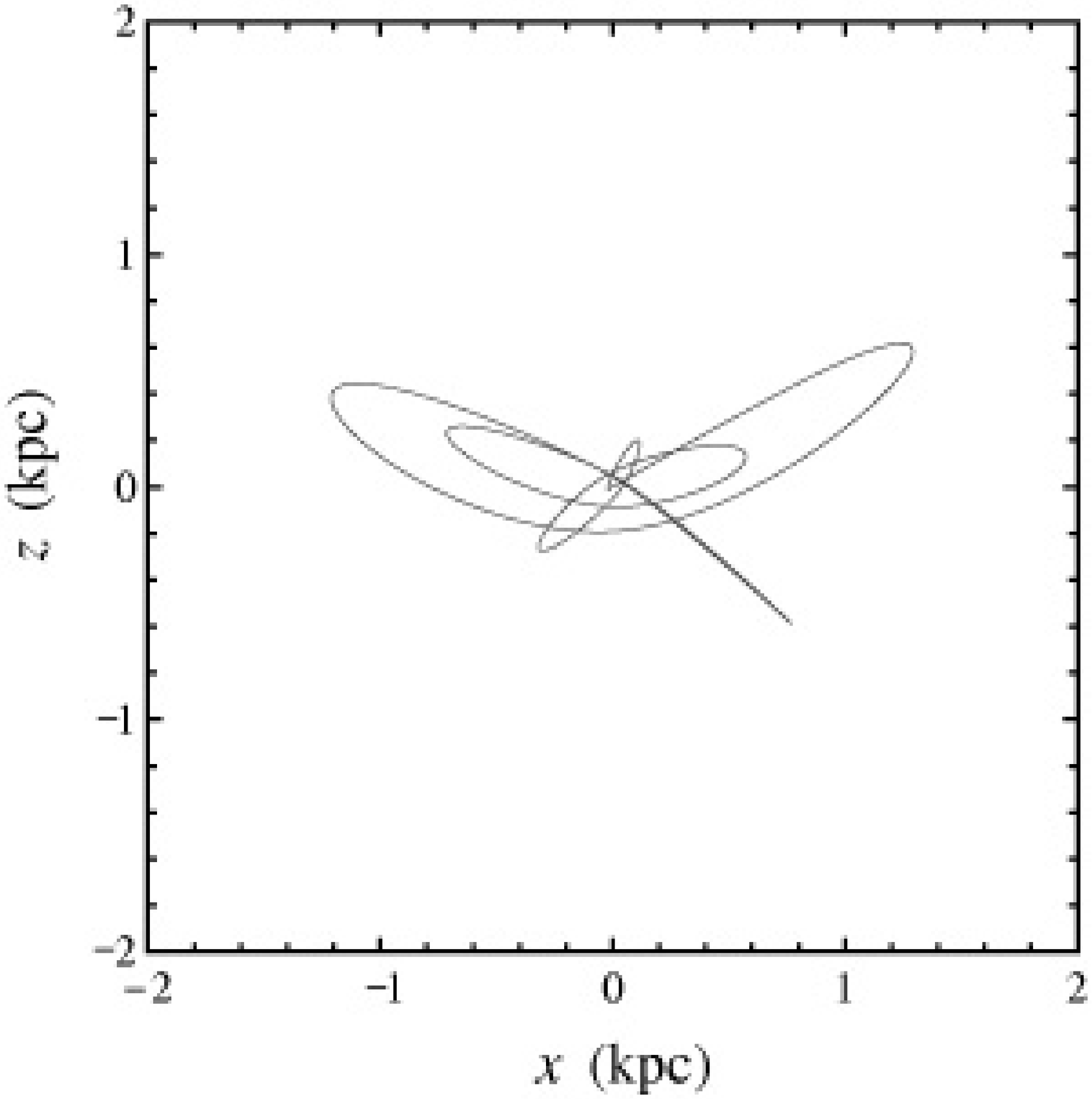} \caption{Same as Figure~\ref{fig:orbitb0}
but for model~d3. \label{fig:orbitd3}}
\end{figure}

For comparison, we also investigate a model with a smaller initial
velocity (model~b0), because we consider a Milky-Way type galaxy, which
is generally expected to experience minor mergers rather than major
mergers. In such cases, large recoil velocities as adopted above would
not be common. A model with a larger SMBH mass is also considered
(model~d3). Figures~\ref{fig:orbitb0} and~\ref{fig:orbitd3} show the
trajectories of the SMBHs for models~b0 and d3, respectively. Their
ejection angles are $\theta=60\arcdeg$. In these models, $v_0$ is too
small (model~b0), or $m_{\rm BH}$ is too large (model~d3) for the SMBH
to be ejected from the spheroidal component of the galaxy. Thus, it
would be difficult to recognize them as recoiling SMBHs, if their host
galaxies are moderately distant. The SMBHs are almost confined to the
$x$-$z$ plane, because the dynamical friction from the spheroidal
component overwhelms that from the galactic disk. Table~\ref{tab:par}
shows that $P_{3e39}$ for models~b0 and d3 is relatively large because
of small $v_0$ and large $m_{\rm BH}$, respectively. The SMBHs set back
to the galactic center in only several orbital periods ($t_{\rm df}\sim
0.01$ and~0.07~Gyr, respectively).

Since we have included the effect of dynamical friction when we consider
the evolution of $L_X$, we can constrain the probability to find SMBHs
with $L_X>L_{\rm th}$ more precisely than Paper~I. It has been estimated
that for comparable mass binaries with dimensionless spin values of 0.9,
only $\sim 10$\% of all mergers are expected to result in an ejection
speed of $\sim 500$--$800\rm\: km\: s^{-1}$ \citep{sch07,bak08}. Since
the ejection speed is smaller for mergers with large mass ratios and
smaller spin values, the actual fraction would be smaller. Although we
consider the mergers of black holes with the masses currently observed
at the centers of disk galaxies, it is unlikely that a galaxy would have
undergone many mergers of black holes with such masses
\citep[e.g.][]{eno04,mic07}. The number of such mergers that a galaxy
has undergone would be $N\la 1$. Thus, since $\tilde{P}_{\rm 3e39}$,
$\tilde{P}_{\rm 1,3e39}\la 0.1$ (Table~\ref{tab:par}), the probability
that a disk galaxy has a traveling SMBH with a luminosity comparable to
or larger than that of ULXs is $\la 1\times 10^{-2}$.

Since the probability is not so large, extensive surveys would be
required to find the SMBHs running in the galactic disks. In the future,
it would be interesting to study whether the probability is larger than
that of finding SMBHs immediately after the ejection from the galactic
centers with velocities of $>1000\rm\: km\:
s^{-1}$\citep[e.g.][]{loe07}. Since the SMBHs ejected into galactic
disks are very bright (Figure~\ref{fig:C3_90}b), they could be observed
even in distant galaxies. As was discussed in Paper~I, observations in
bands other than X-rays would also be useful to detect the SMBHs
orbiting in disk galaxies and discriminate them from IMBHs.

\section{Conclusion}
\label{sec:conclusion}

We have investigated the trajectory of a SMBH ejected from the galactic
center through the emission of gravitational waves at the merger of two
black holes. We included the effect of dynamical friction. For a disk
galaxy comparable to the Galaxy, the orbit decays on a time-scale of
$\ga 10^8$~yr if the initial velocity of the SMBH is $\sim
500$--$800\rm\: km\: s^{-1}$ and the mass is $\sim 10^7\: M_\sun$. The
SMBH accretes the surrounding ISM when it passes the galactic
disk. Since the accretion rate is larger when the relative velocity
between the SMBH and the ISM is smaller, the accretion rate is the
largest when the SMBH passes the apocenter of its orbit that reside in
the galactic disk. Assuming that the accretion flow is a RIAF, we
estimated the X-ray luminosity of the SMBH. We found that the X-ray
luminosity can reach $L_X\ga 3\times 10^{39}\rm\: erg\: s^{-1}$, which
is comparable to or even larger than those of ULXs. In particular, the
X-ray luminosity would reach $L_X\ga 10^{46}\rm\: erg\: s^{-1}$, if the
SMBH is ejected into the galactic plane. Since the probability of
finding the traveling SMBHs with $L_X\ga 3\times 10^{39}\rm\: erg\:
s^{-1}$ in a disk galaxy is $\la 0.01$, extensive surveys would be
required to find them.

\acknowledgments

Y.F. was supported in part by Grants-in-Aid from the Ministry of
Education, Culture, Sports, Science, and Technology of Japan (20540269).

\clearpage

\begin{deluxetable}{cccccccccccc}
\tablewidth{0pt}
\rotate
\tablecaption{Model Parameters and Results\label{tab:par}}
\tablehead{
\colhead{Models} &
\colhead{$m_{\rm BH}$} &
\colhead{$v_0$} &
\colhead{$\ln\Lambda_{\rm disk}$} &
\colhead{$f_{\rm ISM}$} &
\colhead{$r_{\rm max,60}$} &
\colhead{$L_{\rm max,60}$} &
\colhead{$\langle t_{\rm df}\rangle$} &
\colhead{$P_{\rm 3e39}$} &
\colhead{$\tilde{P}_{\rm 3e39}$} &
\colhead{$P_{\rm 1,3e39}$} &
\colhead{$\tilde{P}_{\rm 1,3e39}$} \\
\colhead{} &
\colhead{($M_\sun$)} &
\colhead{($\rm km\: s^{-1}$)} &
\colhead{} &
\colhead{} &
\colhead{(kpc)} &
\colhead{($\rm erg\: s^{-1}$)} &
\colhead{(Gyr)} &
\colhead{} &
\colhead{} &
\colhead{} &
\colhead{} 
}
\startdata
A1    &$3\times 10^6$&500&2.5&0.2&  1 & $1\times 10^{38}$ & 0.63& 0.0058&0.0003& 0     & 0    \\
A2    &$3\times 10^6$&600&2.5&0.2&  2 & $5\times 10^{37}$ & 6.4 & 0.0018&0.0010& 0     & 0    \\
B1    &$1\times 10^7$&500&2.5&0.2&0.7 & $4\times 10^{39}$ & 0.13& 0.23  &0.0027& 0     & 0    \\
B2    &$1\times 10^7$&600&2.5&0.2&  2 & $1\times 10^{39}$ & 1.3 & 0.0756&0.0080& 0.033 &0.0033\\
B3    &$1\times 10^7$&700&2.5&0.2& 10 & $1\times 10^{38}$ &$>$10& 0.031 &0.031 & 0.031 &0.031 \\
C2    &$3\times 10^7$&600&2.5&0.2&  1 & $2\times 10^{40}$ & 0.16& 0.58  &0.0093& 0.10  &0.0016\\
C3    &$3\times 10^7$&700&2.5&0.2&  7 & $1\times 10^{40}$ & 3.7 & 0.31  &0.15  & 0.22  &0.11  \\
C4    &$3\times 10^7$&800&2.5&0.2& 40 & $4\times 10^{38}$ &$>$10& 0.0020&0.0020& 0.0018&0.0018\\
C$'$2 &$3\times 10^7$&600&1.5&0.2&  1 & $2\times 10^{40}$ & 0.16& 0.58  &0.0090& 0.10  &0.0016\\
C$'$3 &$3\times 10^7$&700&1.5&0.2&  7 & $1\times 10^{40}$ & 3.5 & 0.18  &0.073 & 0.12  &0.053 \\
C$''$3&$3\times 10^7$&700&2.5&0.1&  7 & $2\times 10^{39}$ & 3.7 & 0.24  &0.12 & 0.16  &0.097 \\
b0    &$1\times 10^7$&400&2.5&0.2&0.2 & $9\times 10^{39}$ &0.012& 0.83  &0.0010& 0     & 0    \\
d3    &$1\times 10^8$&700&2.5&0.2&  1 & $3\times 10^{41}$ &0.066& 0.95  &0.0062& 0.21  & 0.0013 \\
\enddata
\end{deluxetable}

\clearpage

\clearpage

\end{document}